\documentclass[amsmath,amssymb,aps,twocolumn,showkeys,showpacs]{revtex4}
\usepackage[colorlinks=true, pdfstartview=FitV, linkcolor=blue, citecolor=red, urlcolor=magenta]{hyperref}
\usepackage{graphicx}
\usepackage{latexsym}
\usepackage{amsmath}
\usepackage{amsfonts}
\usepackage{amssymb}
\usepackage{verbatim}
\usepackage{dcolumn}
\usepackage{amssymb}
\usepackage{bm}
\usepackage{float}
\usepackage{subfigure}
\usepackage[english]{babel}
\newcommand{\be}{\begin{equation}}
\newcommand{\ee}{\end{equation}}
\newcommand{\bea}{\begin{eqnarray}}
\newcommand{\eea}{\end{eqnarray}}

\begin{document}

\newcommand{\NITK}{
\affiliation{Department of Physics, National Institute of Technology Karnataka, Surathkal  575 025, India}
}

\newcommand{\IIT}{\affiliation{
Department of Physics, Indian Institute of Technology, Ropar, Rupnagar, Punjab 140 001, India
}}

\title{Ruppeiner Geometry, Reentrant Phase transition and Microstructure of Born-Infeld AdS Black Hole}

\author{Naveena Kumara A.}
\email{naviphysics@gmail.com}
\NITK
\author{Ahmed Rizwan C.L.}
\email{ahmedrizwancl@gmail.com}
\NITK
\author{Kartheek Hegde}
\email{hegde.kartheek@gmail.com}
\NITK
\author{Md Sabir Ali}
\email{alimd.sabir3@gmail.com}
\IIT
\author{Ajith K.M.}
\email{ajith@nitk.ac.in}
\NITK

\begin{abstract}
Born-Infeld AdS black hole exhibits a reentrant phase transition for certain values of the Born-Infeld parameter $b$. This behaviour is an additional feature compared to the van der Waals like phase transition observed in charged AdS black holes. Therefore, it is worth observing the underlying microscopic origin of this reentrant phase transition. Depending on the value of the parameter $b$, the black hole system has four different cases: no phase transition, a reentrant phase transition with two scenarios, or a van der Waals-like (standard) phase transition. In this article, by employing a novel Ruppeiner geometry method in the parameter space of temperature and volume, we investigate the microstructure of Born-Infeld AdS black hole via the phase transition study, which includes standard and reentrant phase transition. We find that the microstructures of the black hole that lead to standard and reentrant phase transitions are distinct in nature. The standard phase transition is characterised by the typical RN-AdS microstructure. In this case, the small black hole phase has a dominant repulsive interaction for the low temperature case. Interestingly, during the reentrant phase transition, displayed by the system in a range of pressures for specific $b$ values, 
the dominant attractive nature of interaction in the microstructure is preserved. Our results suggest that in the reentrant phase transition case, the intermediate black holes behave like a bosonic gas, and in the standard phase transition case the small black holes behave like a quantum anyon gas. In both cases, the large black hole phase displays an interaction similar to the bosonic gas. The critical phenomenon is observed from the curvature scalar, including the signature of the reentrant phase transition.

\end{abstract}


\maketitle


\section{Introduction}

In physics, thermodynamics is a general and powerful tool to understand the physical properties of a system and a wide range of phenomena. The simplicity of thermodynamics lies in its inherent property that, the microscopic details are not necessary to explore the macroscopic physics. This property is particularly useful in dealing with the systems for which the microscopic details are not well understood, such as, quantum gravity. Therefore the lack of full knowledge of quantum aspects of black hole offers a perfect scenario for the application of thermodynamics to probe its microscopic details.

Black holes being thermodynamic systems \citep{Hawking:1974sw, Bekenstein1972, Bekenstein1973, Bardeen1973}, exhibit a rich class of phase transitions, which are the key tools in probing their properties in black hole chemistry. The quite interesting facet of black hole thermodynamics is the phase transition and related phenomenon in AdS spaces. In recent times the interest on AdS black hole thermodynamics aroused among the researchers, after the identification of the cosmological constant with the thermodynamic pressure and the modification in the first law by including the corresponding variations \cite{Kastor:2009wy, Dolan:2011xt}. With this association, it was demonstrated that the phase transition features of AdS black holes can be seen as van der Waals like and/or reentrant phase transitions (RPT) \cite{Kubiznak2012, Gunasekaran2012, Kubiznak:2016qmn}.

A reentrant phase transition (RPT) occurs when the system undergoes more than one phase transition when there is a monotonic change in the thermodynamic variable, such that, the initial and final macrostates of the system are the same. In conventional thermodynamic systems, this phenomenon was first observed in the nicotine/water mixture, in a process with a fixed percentage of nicotine and an increase in the temperature, the system exhibited a reentrant phase transition from initial homogeneous mixed state to an intermediate distinct nicotine/water phases and finally to the homogeneous state \citep{hudson1904gegenseitige}. This kind of reentrant phase transition is observed in a variety of physical systems, more commonly in multicomponent fluid systems, gels, ferroelectrics, liquid crystals, binary gases etc. \citep{1994PhR249135N}. This phenomenon is not limited to condensed matter physics, for example, a $(2+1)$ dimensional Dirac oscillator in non-commutative spacetime and magnetic field shows a similar phase transition \citep{2016JPhCS.670a2040P}.

In black hole physics, reentrant phase transition was first observed in four-dimensional Born Infeld AdS black holes \citep{Gunasekaran2012}, where the initial and final phases are large black holes and the intermediate phase is an intermediate black hole. For this case, LBH/IBH/LBH reentrant phase transition takes place when the temperature is decreased monotonically in a certain range of pressure. However, higher dimensional Born Infeld black holes do not show reentrant phase transitions \citep{Zou:2013owa}. Interestingly, rotating black holes in dimensions $d\geq 6$ show reentrant phase transitions \citep{Altamirano:2013ane}. In subsequent studies, the RPT in higher dimensional single spinning and multi spinning Kerr black holes in anti-de Sitter and de Sitter spacetime were investigated \citep{Altamirano:2013uqa, Altamirano:2014tva, Kubiznak:2015bya}. Reentrant phase transitions were also observed in gravity theories consisting of higher-curvature corrections \citep{Frassino:2014pha, Wei:2014hba, Hennigar:2015esa, Sherkatghanad:2014hda, Hennigar:2015wxa}. The reentrant phase transition of Born Infeld black hole was also analysed with a different perspective, wherein the charge of the system was varied, and the cosmological constant (pressure) was kept fixed \citep{Dehyadegari:2017hvd}. Furthermore, the relationship between the RPT and the photon sphere of Born Infeld AdS spacetime has been studied \citep{Xu:2019yub}. With a motivation from the famous saying by Boltzmann, `If you can heat it, it has microstructure', it is reasonable to ask, what is the underlying microstructure that leads to a reentrant phase transition in a black hole?"

It is a well-established notion that the geometrical methods can serve as a tool to understand microscopic interaction in a thermal system. It was Weinhold who constructed the first thermodynamic geometry method, by considering internal energy as the thermodynamic potential \citep{Weinhold75}. Later, by choosing entropy as the thermodynamic potential, another nifty geometrical method was introduced by Ruppeiner, starting from Boltzmann entropy formula \citep{Ruppeiner95}. Essentially, the geometrical methods were developed from Gaussian thermodynamic fluctuation theory, in which a metric is constructed by choosing a suitable thermodynamic potential in a phase space which constitutes other thermodynamic variables, the corresponding curvature scalar encodes details about phase transitions and critical points. This method was used in analysing the conventional thermal systems like ideal fluids, van der Waals (vdW) systems, Ising models, quantum gases etc. \citep{Ruppeiner79, Janyszek1990b, Ruppeiner81, Janyszek89, Janyszek_1990, Oshima_1999x, Mirza2008, May2013}. The results so obtained give a very clear picture of the applicability of the geometrical methods in understanding microscopic details. The two main aspects of Ruppeiner geometry is the revelation of correlation and the interaction type of the microstructure. The sign of the curvature scalar is an indicator of the nature of the interaction, positive/negative for repulsive/attractive interaction and zero for no interaction. On the other hand, the magnitude of the curvature scalar is the measure of correlation length of the system. As the correlation length diverges near the critical point of the system, so does the curvature scalar.

As the entropy is taken to be the thermodynamic potential in Ruppeiner geometry, the application of it to black hole system is straight forward and interesting (see Ref. \citep{Ruppeiner:2013yca} for a recent review on this). An early accomplishment of this method is to the BTZ black holes \citep{Cai:1998ep}. Later, the nature of microstructure of Reissner-Nordstrom (RN) black hole was sought for using this method \citep{Aman:2003ug}, where the curvature scalar $R$ vanishes and no interaction was found. However, a non-vanishing $R$ was expected as the spacetime is curved. This problem was re-examined by considering a complete set of thermodynamic variables to begin with, including the angular momentum and cosmological constant, and taking appropriate limit for RN black hole \citep{Mirza:2007ev}, which lead to a non-vanishing $R$. In subsequent developments, the vdW like behaviour of the black hole and the underlying microscopic details were investigated using the geometrical methods \citep{Sahay:2010wi, Chaturvedi:2014vpa, Wei:2017icx, Chaturvedi:2017vgq}. In this so-called $R$-crossing method, the coordinates of the parameter space are taken as temperature and fluid density, and the diverging behaviour of $R$ can be observed.  However mismatch in the divergence of curvature scalar and specific heat in some cases using this method, lead to the proposal of several other geometrical methods \citep{Quevedo:2006xk, Liu:2010sz,Niu:2011tb,Wei:2012ui,Banerjee:2011cz,Mansoori:2013pna,Mo:2013sxa,Mansoori:2014oia, Hendi:2015cka,Dolan:2015xta,Mansoori:2016jer,HosseiniMansoori:2019jcs,Banerjee:2010da,Banerjee:2010bx}. 

Combining the idea of the dynamic cosmological constant in black hole chemistry with the thermodynamic geometry, a new method of investigating the black hole microstructure was proposed by Wei et. al. \citep{Wei2015}. In this method, mass and pressure were taken as the coordinates the parameter space, and a new concept of the number density of black hole molecules was introduced. The most spectacular outcome of this construction was the existence of a repulsive interaction in the microstructure of a charged AdS black hole. Soon, this method was adopted to investigate various aspects of black hole microstructure by several researchers \citep{Guo2019, Du2019, Dehyadegari2017, Chabab2018,  Deng2017,  Zangeneh2017, Miao2017, Miao2019a, Chen2019, Xu:2019nnp,  Kumara:2019xgt, Kumara:2020mvo}. Despite the substantial generalisations and applications of this method, the scalar curvature thus constructed was lacking the characteristic divergence behaviour near the critical point. This imperfection compared to the earlier geometrical methods called for a reanalysis of the basic setting of the methodology. 

Recently, a new revised method to rectify the above shortcoming was proposed by Wei et. al. within the framework of  Ruppeiner geometry \citep{Wei2019a, Wei2019b}. The key issue lurking in the previous attempts were the non-independence of the thermodynamic variables entropy and volume for a spherically symmetric AdS black hole. This leads to a vanishing specific heat, and hence a singularity in line element and a divergent curvature scalar. In the new approach, treating the specific heat as a tiny constant close to zero, and employing the temperature and volume as fluctuation coordinates, a normalised curvature scalar was defined to probe the black hole microstructure. The new curvature scalar aptly features the critical phenomena and microstructure interactions of the black hole with universal properties. Therefore this novel method promptly captured much attention \citep{Wei:2019ctz, Kumara:2020ucr, Wei:2020poh, Wu:2020fij, Xu:2019gqm, Ghosh:2019pwy, Ghosh:2020kba, Yerra:2020oph, Dehyadegari:2020ebz}. In this article, we aim to understand the properties of the underlying microstructure that leads to reentrant phase transition in a Born-Infeld AdS black hole using this novel method.

It is worth noting that the investigation of the microstructure of the black hole system has been one of the major challenges in black hole physics for the past few decades. Even though string theory and loop quantum gravity provide tools to understand quantum gravity, there is no complete theoretical description for it. Therefore the microscopic physics of black holes is confined to some phenomenological approaches. The success of black hole thermodynamics in understanding various aspects of the black hole prompts us to formulate some relation between micro-dynamics and thermodynamics. However, the process is in a reverse sense compared to statistical physics, where the microscopic knowledge is sought from macroscopic details. However, when we seek the nature of microstructure, we do not have a clear picture of the black hole constituents, we take an abstract concept that black holes are constituted of black hole molecules.

This article is organised as follows. In the next section, we discuss the thermodynamics and the phase structure of the black hole. We will consider the van der Waals case (which will be called SPT throughout the article which stands for standard phase transition) and RPT case separately. Then the microstructure study is carried out by constructing the Ruppeiner geometry using the fluctuation coordinates as temperature and volume  (section \ref{sectwo}). In the final section (\ref{secthree}) we present our results.


\section{Thermodynamics and Phase Structure of the Black Hole}
\label{secone}
In this section we present thermodynamics and phase structure of the Born-Infeld AdS (BI-AdS) black hole. The action for Einstein gravity in the presence of Born-Infeld field has the following form \citep{Born:1934gh, Gunasekaran2012},
\begin{equation}
S=\frac{1}{16\pi}\int d^4 x\sqrt{-g}\left[\mathcal{R}-2\Lambda+4 b^2\left(1-\sqrt{1+\frac{F_{\mu\nu}F^{\mu\nu}}{2b^2}}\right)\right].
\end{equation}
Here $\mathcal{R}$ and $\Lambda$ are the Ricci scalar and cosmological constant, respectively. The Born Infeld parameter $b$  with dimension of mass represents  the maximal electromagnetic field strength, and is related to the string tension (the identification is motivated from string theory \citep{Gibbons:2001gy}), and $F_{\mu\nu}=\partial_\mu A_\nu-\partial_\nu A_\mu $ is the electromagnetic field tensor, where $A_\mu$ is the vector potential. We study the extended thermodynamics of the black hole, where the cosmological constant serves the role of thermodynamic pressure. Their relation to the AdS radius $l$ are given by,
\begin{equation}
\Lambda=-\frac{3}{l^2}, \qquad \text{and} \qquad P=\frac{3}{8\pi l^2}.
\end{equation}
The solution to the Einstein field equations in the static and spherically symmetric spacetime background yields \citep{Fernando:2003tz, Dey:2004yt, Cai:2004eh},
\begin{equation}
ds^2=-f(r) dt^2+ f^{-1}(r) dr^2 + r^2\left(d\theta^2 +\sin^2\theta d\phi^2 \right)
\end{equation}
\begin{equation}
F=\frac{Q}{\sqrt{r^4+ Q^2/b^2}}dt \wedge dr,
\end{equation}
with the metric function of the form,
\begin{align}
f(r)=&1+\frac{r^2}{l^2}-\frac{2M}{r}+\frac{2b^2}{r}\int_r^{\infty}\left(\sqrt{r^4+\frac{Q^2}{b^2}-r^2}\right)dr\\
   =& 1+\frac{r^2}{l^2}-\frac{2M}{r}+\frac{2b^2r^2}{3}\left(1-\sqrt{1+\frac{Q^2}{b^2r^4}}\right)\nonumber \\
   &+\frac{4Q^2}{3 r^2}{}_2F_1\left(\frac{1}{4},\frac{1}{2},\frac{5}{4};-\frac{Q^2}{b^2r^4} \right).
\end{align}
Where, $_2F_1$ is the hypergeometric function, the parameters $M$ and $Q$ are the ADM mass and the asymptotic charge of the solution, respectively. We obtain mass of the black hole by setting $f(r_+)=0$,
\begin{align}
M=& \frac{r_+}{2}+\frac{r_+^3}{2l^2}+\frac{b^2 r_+^3}{3}
\left(1-\sqrt{1+\frac{Q^2}{b^2 r_+^4}}\right)\nonumber\\
&+\frac{2Q^2}{3 r_+^2}{}_2F_1\left(\frac{1}{4},\frac{1}{2},\frac{5}{4};-\frac{Q^2}{b^2 r_+^4} \right).
\end{align}
The Hawking temperature and the corresponding entropy are given by,
\begin{align}
T=&\frac{1}{4\pi r_+}\left( 1+\frac{3 r_+}{l^2} + 2b^2 r_+^2
\left(1-\sqrt{1+\frac{Q^2}{b^2 r_+^4}}\right)\right),\\
S=&\frac{A}{4}=\pi r_+^2.
\end{align}
The electric potential $\Phi$ and the electric polarization $B$, which is conjugate to $b$ and is referred to as “BI vacuum polarization”, measured at infinity with respect to the event horizon they are calculated to be,
\begin{align}
\Phi=&\frac{Q}{r_+}{}_2F_1\left(\frac{1}{4},\frac{1}{2},\frac{5}{4};-\frac{Q^2}{b^2r_+^4} \right)\\
B=&\frac{2}{3}br_+^3\left(1-\sqrt{1+\frac{Q^2}{b^2r_+^4}}+\frac{Q^2}{3br_+}\right){}_2F_1\left(\frac{1}{4},\frac{1}{2},\frac{5}{4};-\frac{Q^2}{b^2r_+^4} \right)
\end{align}
Interpreting the mass $M$, as enthalpy rather than the internal energy of the black hole, we obtain the first law of thermodynamics as,
\begin{equation}
 dM= TdS+ VdP +\Phi dQ + B db,
\end{equation}
where $V=\frac{4\pi r_+^3}{3}$ is the thermodynamic volume of the system. In addition to the first law of thermodynamics, the thermodynamic quantities of BI-AdS black hole satisfy the Smarr formula, which is obtained by scaling argument as,
\begin{eqnarray}
M= 2\left(TS-VP\right)+ \Phi Q - Bb.
\end{eqnarray}
The equation of state of the black hole system is,
 \begin{equation}
 P=\frac{T}{2 r_+}-\frac{1}{8\pi r_+^2}-\frac{b^2}{4\pi}\left(1-\sqrt{1+\frac{ Q^2}{b^2 r_+^4}}\right).
 \end{equation}
The black hole shows a van der Waals like phase transition, which depends on the value of the Born Infeld coupling coefficient $b$. The critical values corresponding to this phase transition is obtained by employing the conditions,
\begin{equation}
 \left( \partial_{r_+} P\right)_T= \left( \partial_{r_+,r_+} P\right)_T=0.
 \end{equation}
The critical values are \citep{Gunasekaran2012, Xu:2019yub},
\begin{align}
P_c=&\frac{1-16x Q^2}{8\pi {r_+^2}_c}-\frac{b^2}{4\pi}\left(1-\frac{1}{4x{r_+^2}_c}\right), \\
T_c=& \frac{1-8 x Q^2}{2 \pi {r_+}_c},\\
{r_c}_+=& \frac{1}{2}\left( \frac{1}{x_k^2}-\frac{16 Q^2}{b^2}\right)^{\frac{1}{4}},
\end{align}

where
\begin{align}
 x_k=2\sqrt{-\frac{p}{3}}\cos\left( \frac{1}{3} \arccos\left( \frac{3q}{2p}\sqrt{\frac{-3}{p}}\right)-\frac{2\pi k}{3}\right),\\
 k=0,1,3, \nonumber 
\end{align}
and
\begin{equation}
p=-\frac{3b^2}{32 Q^2},\qquad q= \frac{b^2}{256 Q^4}.
\end{equation}
Since, the critical point corresponding to $x_2$ is always a complex number, effectively we have only two critical points. Based on these two values we can classify the system into four different cases which depend on the value of $b$. 

\begin{itemize}
    \item Case 1 (no PT), $b<b_0$:
    For this condition, the system behaves like a Schwarzschild AdS black hole. The large black hole phase is stable and the small black hole phase is unstable. Therefore there is no van der Waals like phase transition in this case.
   \item Case 2 (RPT), $b_0<b<b_1$:
   In this condition, the system is characterised by two critical points $c_0$ and $c_1$. However $c_0$ is an unstable point as it has a higher Gibbs free energy. In this scenario, the system exhibits a zeroth-order phase transition and a van der Waals like first-order phase transition,  between a large black hole phase and an intermediate black hole phase. These successive transitions are together termed as a reentrant phase transition.
\item Case 3 (RPT), $b_1<b<b_2$: This is another case of reentrant phase transition displayed by the black hole. However, here the critical point $c_0$ lies in the negative pressure region. 

\item Case 4 (SPT), $b_2<b$: Here the black hole exhibits the typical van der Waals like phase transition with one critical point. 
\end{itemize}

The values of the parameter $b$ where the phase transition behaviour changes, are given by,

\begin{align}
b_0=&\frac{1}{\sqrt{8}Q}\approx \frac{0.3536}{Q}~,~~b_1=\frac{\sqrt{3+2\sqrt{3}}}{6 Q}\approx\frac{0.4237}{Q},\nonumber \\
b_2=&\frac{1}{2 Q}= \frac{0.5}{Q}.
\end{align}
For SPT situation, we introduce the reduced parameters, which are defined as,
\begin{equation}
    P_r=\frac{P}{P_c}\qquad T_r=\frac{T}{T_c} \qquad V_r=\frac{V}{V_c}.
    \label{red1}
\end{equation}

\begin{figure*}[t]
\centering
\subfigure[][]{\includegraphics[scale=0.9]{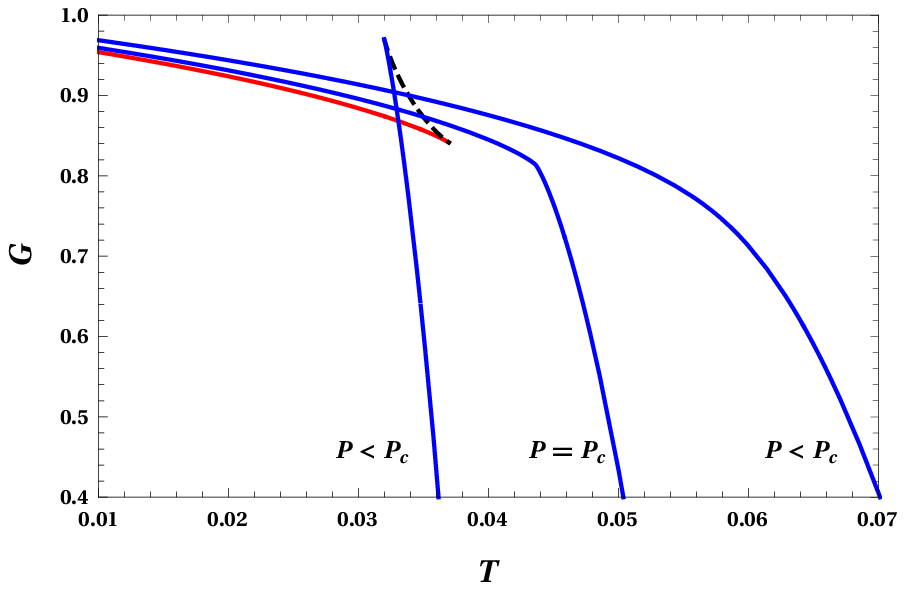}\label{SPTgt1}}
\qquad
\subfigure[][]{\includegraphics[scale=0.9]{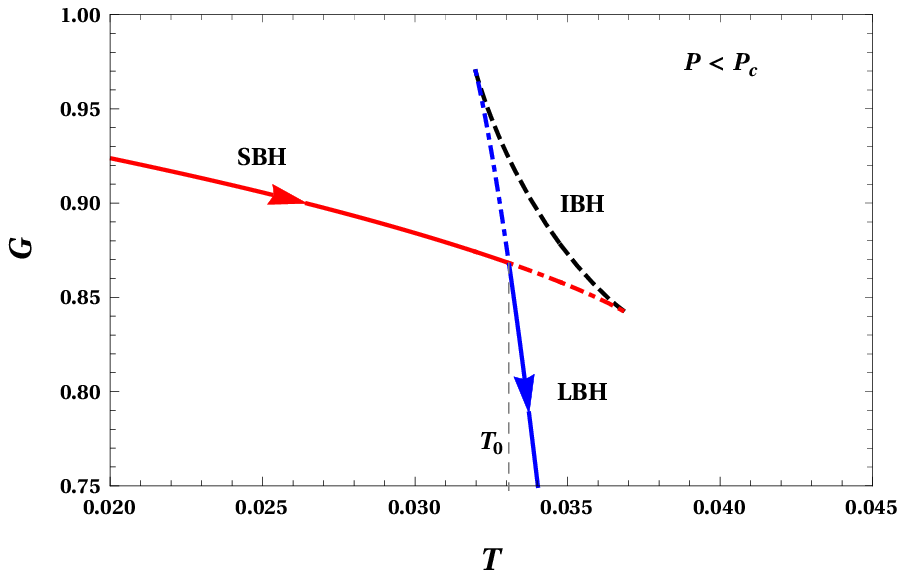}\label{SPTgt2}}
\caption{The behaviour of the Gibbs free energy $G$ for the SPT case. Here, we take $Q=1$ and $b=1$. In Fig. \ref{SPTgt1}, the blue and red solid lines stand for stable states, whereas black dashed line for unstable states. In Fig. \ref{SPTgt2} the direction of evolution of the system with increasing temperature $T$ is depicted. The solid red and blue lines correspond to SBH and LBH phase of the black hole, respectively. The dot-dashed lines are the stable states not followed by the system.}
\label{SPTgt}
\end{figure*}

\begin{figure*}[t]
\centering
\subfigure[][]{\includegraphics[scale=0.9]{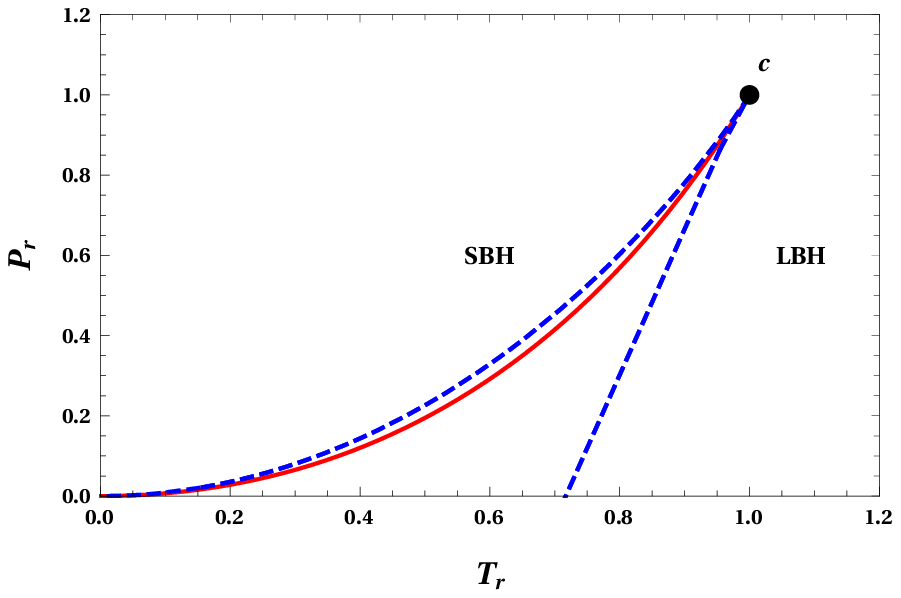}\label{SPTPT}}
\qquad
\subfigure[][]{\includegraphics[scale=0.9]{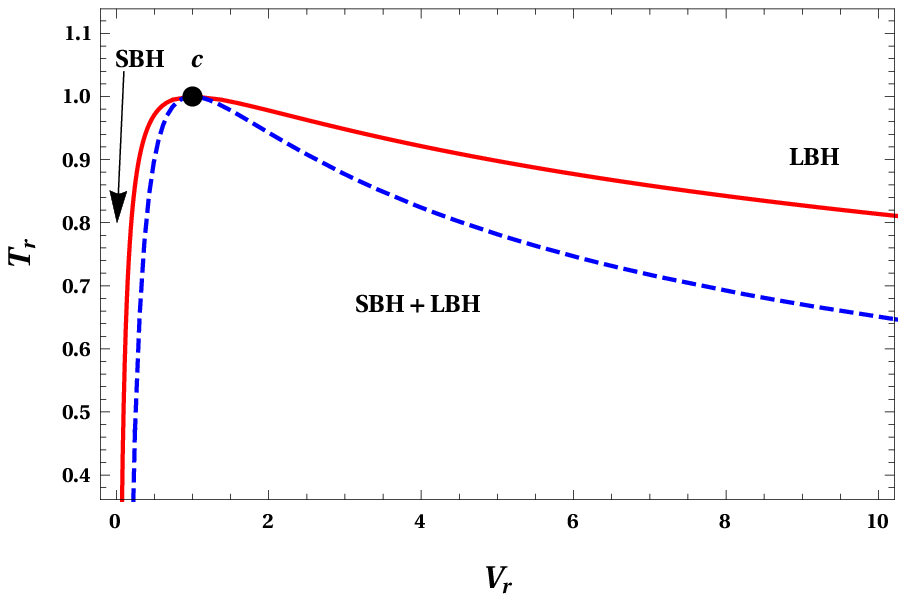}\label{SPTTV}}
\caption{The coexistence curve (red solid line) and the spinodal curve (blue dashed line) for the SPT of the black hole. The coexistence curve separates SBH and LBH phases. The region between the coexistence curve and spinodal curves are the metastable SBH and LBH phases in respective regions. The black dot $c$ denotes the critical point. Here we have taken $Q=1$ and $b=1$. }
\label{SPTphase}
\end{figure*}

\begingroup
\setlength{\tabcolsep}{8pt} 
\renewcommand{\arraystretch}{1.2} 
\begin{center}
\begin{table*}
\centering
\begin{tabular}{ cccccc } 
\hline
\hline
$b$ & 0.6 & 0.7 & 0.8 & 0.9 & 1 \\
\hline
$P_c$ & 0.003486 & 0.003436 & 0.003405 & 0.003385 & 0.003371 \\
$T_c$ & 0.044251  & 0.043981 & 0.043815 & 0.043705 & 0.043628 \\ 
\hline
\hline
\end{tabular}
\caption{\label{sptcritical} The critical values $P_c$ and $T_c$ for different values of $b$, with $Q=1$, corresponding to the SPT of the black hole.}
\end{table*}
\end{center}
\endgroup

Since there are two critical points in the RPT case, there is no unique way of defining the reduced parameters. However, there is only one critical point which appears in the phase transition scenario of the black hole, the point where the first-order transition terminates. We define reduced parameters considering that critical point as,
\begin{equation}
    P_r=\frac{P}{P_{c1}}\qquad T_r=\frac{T}{T_{c1}} \qquad V_r=\frac{V}{V_{c1}}.
    \label{red2}
\end{equation}

A clear picture of phase transition of BI-AdS black hole can be understood from the behaviour of the thermodynamic potential, as it determines the globally stable states of equilibrium thermodynamics. The thermodynamic potential for the system with fixed temperature $T$, pressure $P$ and charge $Q$ is the Gibbs free energy, which is calculated from the Euclidean action \citep{Kubiznak2012, Gunasekaran2012}. The stable state of the system then corresponds to the lowest Gibbs free energy. The expression for the Gibbs free energy is given by,

\begin{align}
G(T,P) = \frac{1}{4}\left[r_+-\frac{8\pi}{3}P r_+^3-\frac{2b^2r_+^2}{3}\left(1-\sqrt{1+\frac{Q^2}{b^2r_+^4}}\right)\right. \nonumber \\
\left. +\frac{8Q^2}{3 r_+}{}_2F_1\left(\frac{1}{4},\frac{1}{2},\frac{5}{4};-\frac{Q^2}{b^2r_+^4} \right)\right].
\end{align}

\subsection{Standard Phase Transition Case}

First, we study the Gibbs free energy for the SPT case by choosing $Q=1$ and $b=0.45$, the result is shown in Fig. \ref{SPTgt}. In these plots, a simple measure used for thermodynamic stability is the positivity of the specific heat. The negativity of specific heat stands for thermodynamic instability. As we are working in the canonical ensemble in an extended space the specific heat under consideration is the specific heat at constant pressure $C_P$. In this case (SPT), the Gibbs free energy exhibits a typical swallowtail behaviour, which is the signature of a first-order phase transition. This behaviour is seen for $P<P_c$, which disappears at $P=P_c$, shown in Fig. \ref{SPTgt1}, which is a second-order phase transition point. The system always prefers a state with a low Gibbs free energy. From Fig. \ref{SPTgt2} it is clear that initially in low-temperature region, the system chooses a SBH phase as it has a lower Gibbs free energy. As the temperature increases, it follows the LBH phase branch at $T_0$, as it has a lower Gibbs free energy. At this point, the system undergoes a first-order transition from SBH to LBH phase. This transition is similar to the liquid-gas phase transition of a vdW fluid.

The phase structure of the black hole for the SPT case is presented in Fig. \ref{SPTphase} using the coexistence and spinodal curves in the reduced parameter space. There exists no analytical expression for the coexistence curve of the system, therefore we obtain it via a numerical method by observing the swallowtail behaviour of Gibbs free energy.  The coexistence curves so obtained in the $P_r-T_r$ plane are inverted for $T_r-V_r$ plane. In the $P_r-T_r$ plane, Fig. \ref{SPTPT}, the coexistence curve separates the SBH and LBH phases of the black hole. The coexistence region of SBH and LBH can be understood from $T_r-V_r$ plane, Fig. \ref{SPTTV}, which is the region under the coexistence curve. In the coexistence region, the equation of state is not applicable. To identify the metastable phases of the black hole, we find the spinodal curves, blue dashed lines in Fig. \ref{SPTphase}, which are defined by,

\begin{equation}
(\partial _V P)_T=0 \qquad \text{and} \qquad (\partial _V T)_P=0.
\label{spinodaleqn}
\end{equation}

The spinodal curves meet the coexistence curve at the critical point. The region between the spinodal curve and coexistence curve corresponds to metastable phases. The region adjacent to the LBH/SBH phase is the metastable LBH/SBH phase. Beyond the critical point $c$, the distinction between the phases is not possible, which is termed as supercritical black hole region. In the $P_r-T_r$ plane the upper spinodal curve begins from zero, whereas lower one starts from $T_r=0.716953$. The latter one is slightly higher than that of RN-AdS black hole, where it is $T_r=\sqrt{2}/2\approx0.707107$, and little less than $27/32\approx0.84375$ of vdW fluid \citep{Wei2019b}. In the $T_r-V_r$ plane, the curve intercepts the $x$-axis at $V_r=0.163239$ for $T_r=0$, which is comparable to $V_r=1/3\sqrt{3}\approx0.19245$ of RN-AdS case \citep{Wei2019b}. However it is not the exact numerical value which is of interest, but the non zero value compared to zero value for the vdW fluid. This non-zero value is related to the repulsive interactions in the microstructure, which we will study in the next section. For completeness, we mention that the spinodal curve adjacent to the LBH phase in $T_r-V_r$ plane approaches infinity. Before concluding the SPT case, we list the critical values of pressure and temperature for different values of $b$ in table \ref{sptcritical}, which we will be using to investigate the microstructure in section \ref{sectwo}.

\begin{figure*}[t]
\centering
\subfigure[][]{\includegraphics[scale=0.9]{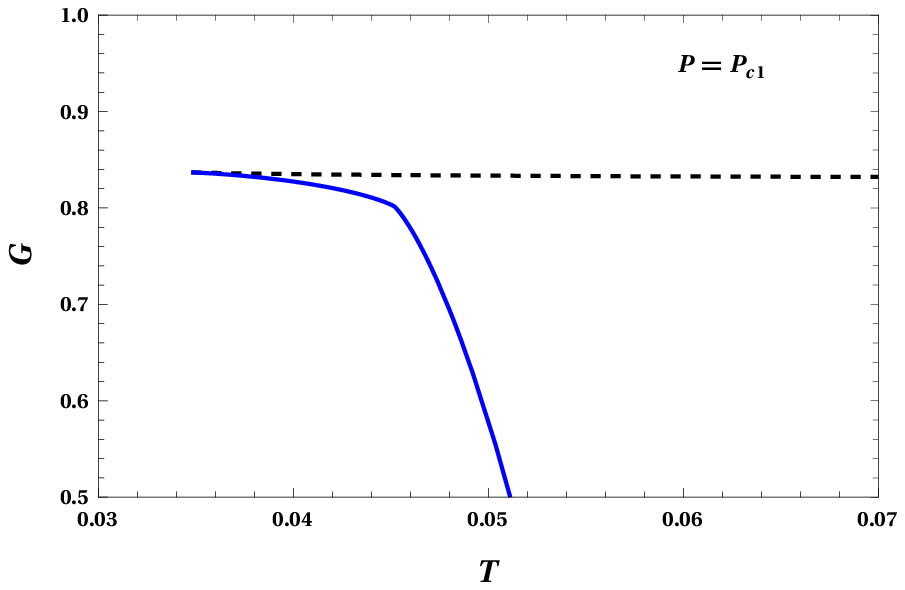}\label{gplot1}}
\qquad
\subfigure[][]{\includegraphics[scale=0.9]{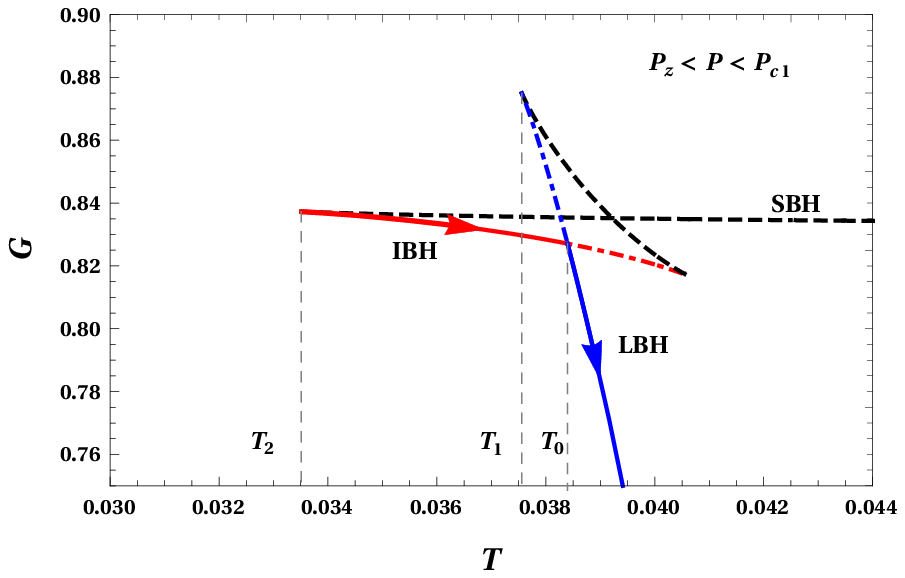}\label{gplot2}}
\qquad
\subfigure[][]{\includegraphics[scale=0.9]{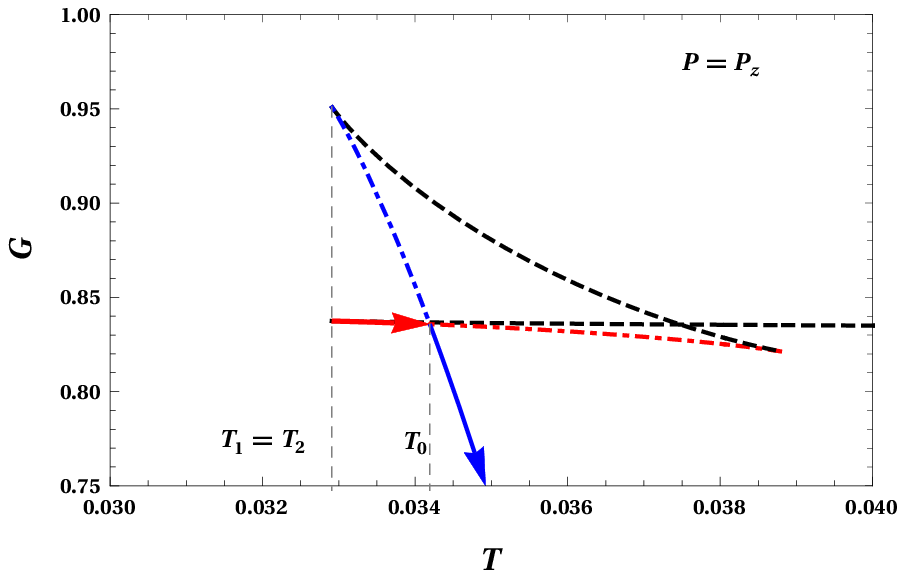}\label{gplot3}}
\qquad
\subfigure[][]{\includegraphics[scale=0.9]{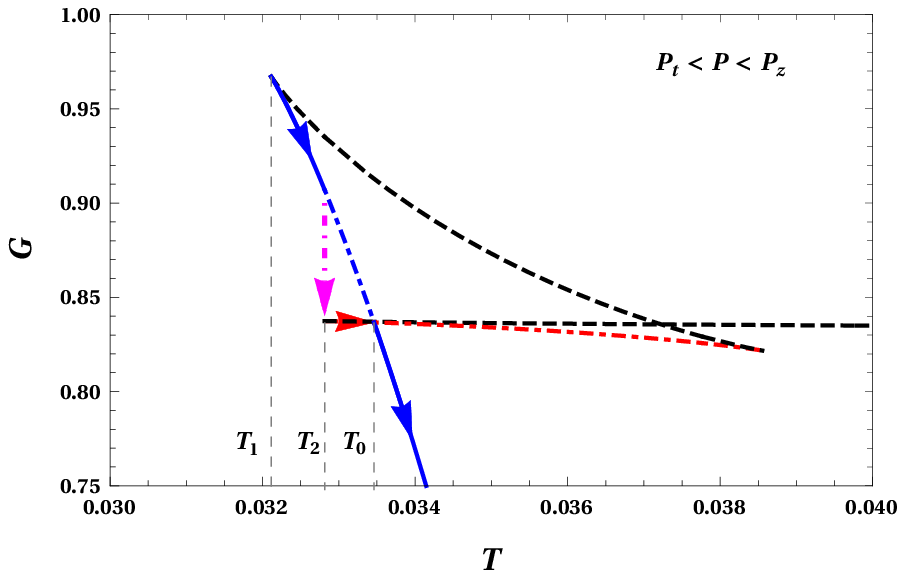}\label{gplot4}}
\qquad
\subfigure[][]{\includegraphics[scale=0.9]{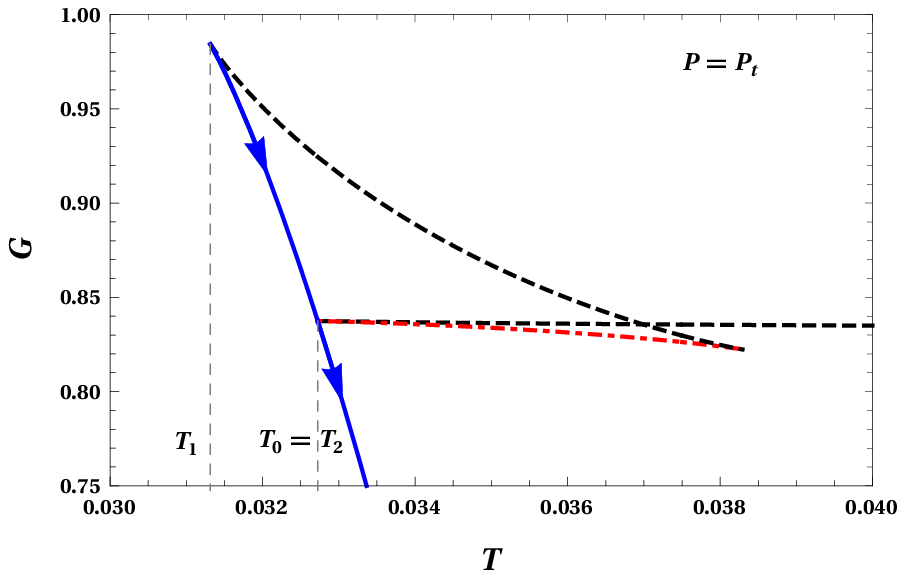}\label{gplot5}}
\qquad
\subfigure[][]{\includegraphics[scale=0.9]{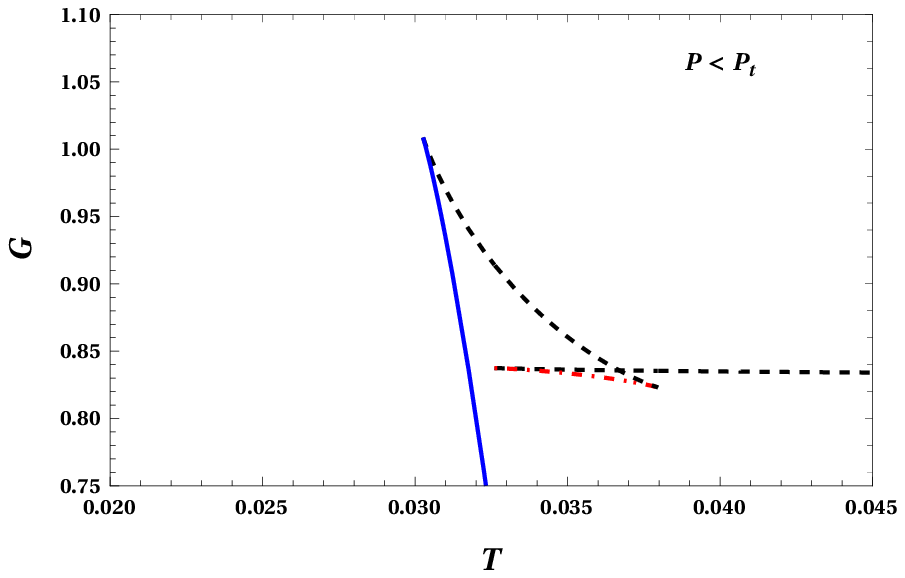}\label{gplot6}}
\caption{The behaviour of the Gibbs free energy $G$ for the RPT case. We take $Q=1$ and $b=0.45$. The black dashed lines correspods to negative $C_P$, whereas blue and red lines correspond to positive $C_P$. The solid red and blue lines are states preferred by the system over the dot-dashed red and blue lines. }
\label{RPTgt}
\end{figure*}

\subsection{Reentrant Phase Transition Case}

\begin{figure*}[t]
\centering
\subfigure[][]{\includegraphics[scale=0.9]{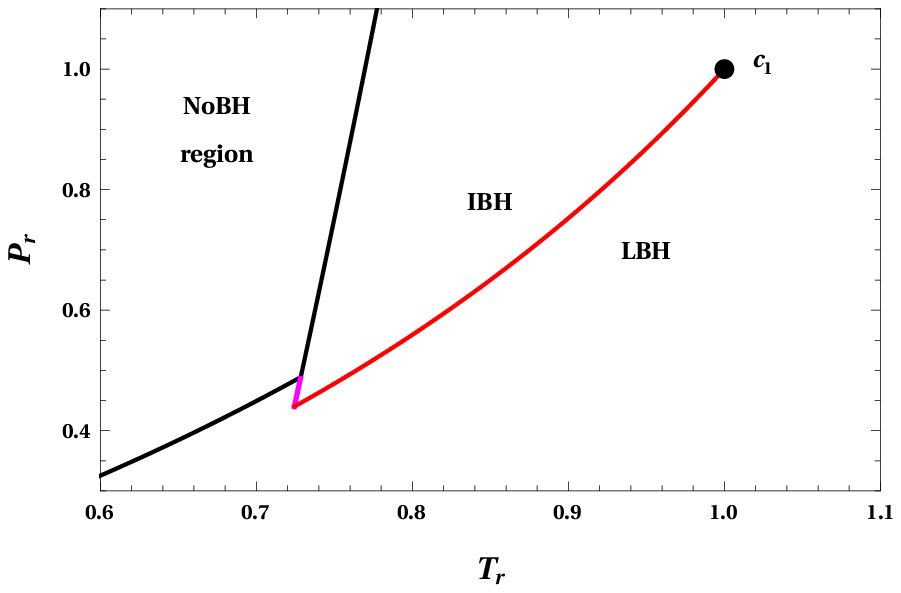}\label{RPTnbh}}
\qquad
\subfigure[][]{\includegraphics[scale=0.9]{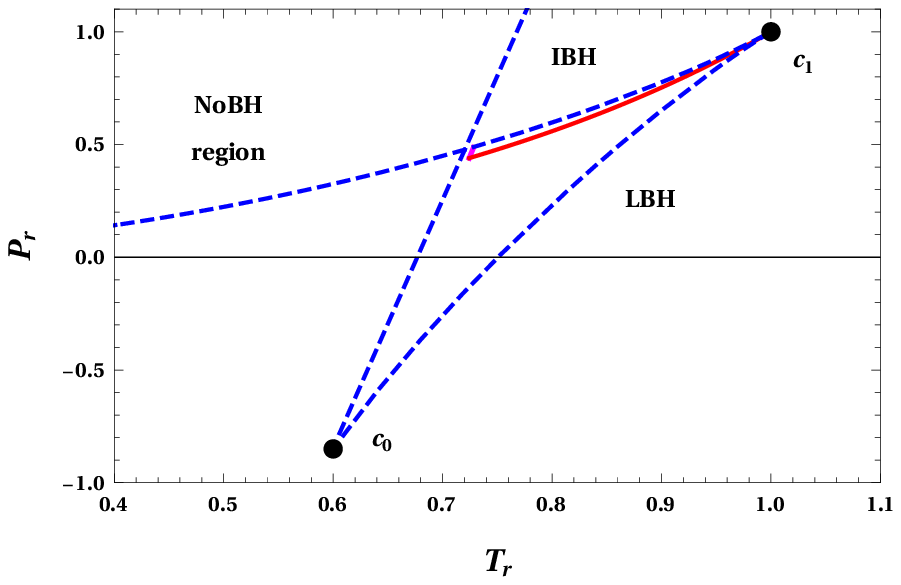}\label{RPTPT}}
\qquad
\subfigure[][]{\includegraphics[scale=0.9]{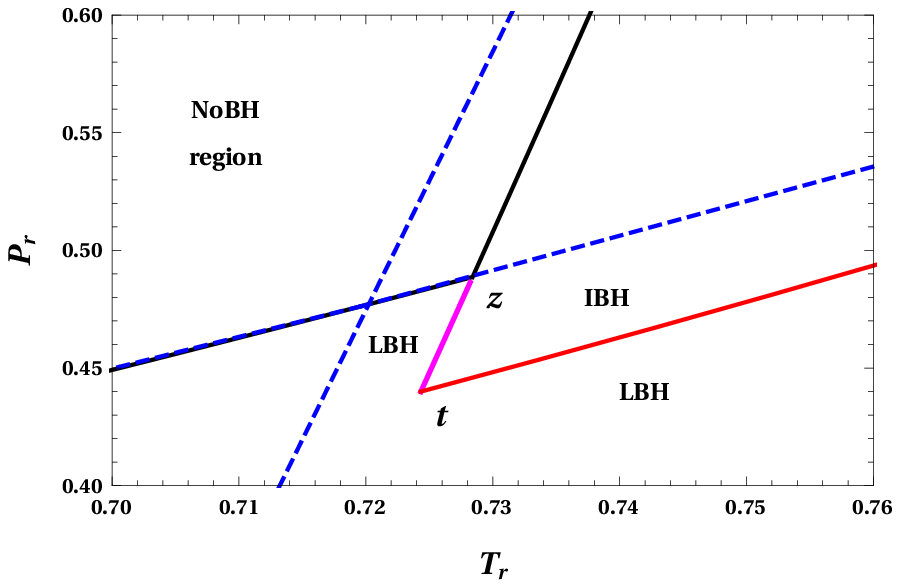}\label{RPTPTsub}}
\qquad
\subfigure[][]{\includegraphics[scale=0.9]{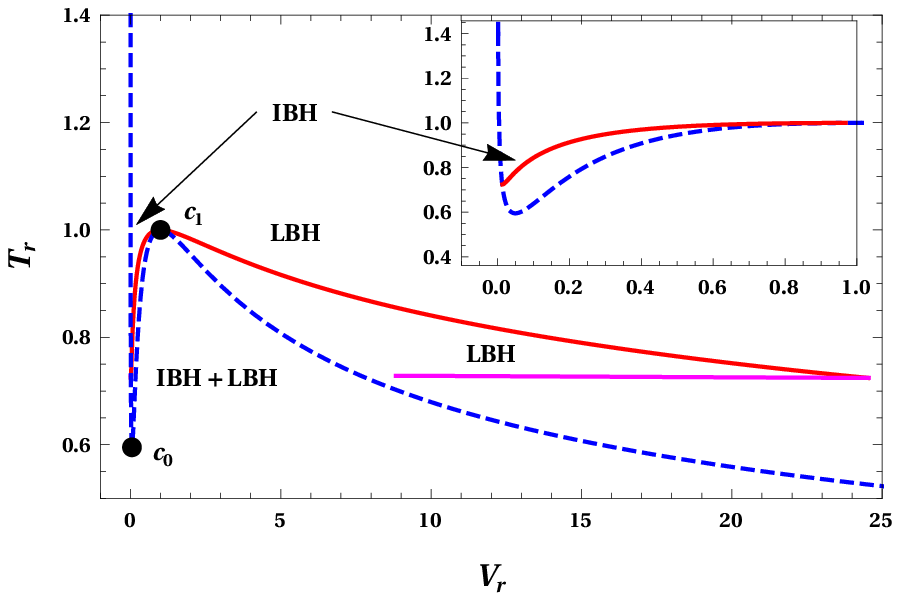}\label{RPTTV}}
\caption{Phase diagrams for reentrant case. (a) Black line separates no black hole region from black hole region. First order coexistence curve (solid red) and zeroth order phase transition line (solid magenta) are also shown. (b) The transition lines and the spinodal curves for the RPT of the black hole. $c_0$ and $c_1$ are the two critical points. At $c_1$ spinodal curve meets first order coexistence curve. (c) The enlarged reentrant phase transition region. The points $t$ and $z$ correspond to $(P_t,T_t)$ and $(P_z,T_z)$, respectively. (d) Phase structure in the $T_r=V_r$ plane. Here, the near origin behaviour is shown in the inset. In all these plots we have taken $Q=1$ and $b=0.45$ }
\label{RPTphase}
\end{figure*}

Now we focus on the reentrant phase transition exhibited by the BI-AdS black hole. Among the two cases of reentrant phase transition (case 2 and case 3), we chose case 3 $(b_1<b<b_2)$ in this article to demonstrate the associated properties. The results obtained are applicable for the case 2 of $b_0<b<b_1$ also. The reentrant phase transition can be better understood from Gibbs free energy study, as shown in Fig. \ref{RPTgt}.  In the $G-T$ plots, for pressure $P=P_{c1}$, which corresponds to the second-order phase transition point, there is no swallow tail behaviour, as seen from Fig. \ref{gplot1}. The solid blue line represents a stable phase of the system (LBH) with positive specific heat at constant pressure $C_P$, while the dashed black line stands for unstable phase (SBH) of the system with negative $C_P$. We emphasise the fact that SBH phase in RPT case is always unstable. There is no phase transition above the critical pressure $P_{c1}$. For pressures $P<P_{c1}$, we begin to observe a first-order phase transition, the indication being the swallowtail, Fig. \ref{gplot2}. Here, we identify three temperatures, designated as $T_0$, $T_1$ and $T_2$. The points in the curve corresponding to $T_1$ and $T_2$ correspondingly connect the stable LBH (blue line) and stable IBH (red line) branches to unstable branches (dashed black). The temperature $T_0$, which is the intersection of the stable IBH and LBH branches, is the point where a vdW like IBH-LBH first-order phase transition occurs. As we decrease the pressure, the mere first-order phase transition situation continues till $P=P_z$, where we have $T_1=T_2\equiv T_z$, which is depicted in Fig \ref{gplot3}. Further decreasing the pressure, $P<P_z$, leads to a scenario where an additional zeroth-order phase transition is also exhibited by the system, Fig \ref{gplot4}. This happens for a range of pressures $P_t<P<P_z$, with $T_1<T_2<T_0$. For a fixed pressure in this range, if the system is at some temperature $T_1$ initially, then it stays in LBH phase as the temperature increases, till $T=T_2$. At temperature $T_2$, the system finds a stable branch with lower Gibbs energy and jumps into the IBH phase. Unlike a first-order phase transition, there is a finite change in Gibbs free energy during this transition. Further, the system undergoes a secondary phase transition at $T=T_0$ from IBH to LBH phase, which is a first-order transition. In effect, the system undergoes LBH/IBH/LBH phase transition, the initial and final phases being the same. This is termed as a reentrant phase transition. This phenomenon disappears at $P=P_t$, with $T_0=T_2\equiv T_t$, where the system has only LBH phase, Fig \ref{gplot5}. There is no zeroth order or vdW like first-order phase transitions possible for the pressures $P<P_t$, Fig. \ref{gplot6}, wherein only the LBH phase exists. In short, the system exhibits both first order and zeroth-order phase transition for the range of pressures $P \in (P_t, P_z)$ and temperatures $T \in (T_t, T_z)$, whereas only first-order phase transition for $P \in (P_z, P_{c1})$ and $T \in (T_z, T_{c1})$.

The phase diagrams of the black hole corresponding to the reentrant phase transition case is presented in Fig. \ref{RPTphase}. As in SPT case, here also the results are obtained numerically by observing the behaviour of Gibbs free energy.  Interestingly, for all values of pressure, we find that there is a lowest temperature below which there is no existence of black hole. The line which separates the black hole solutions from no black hole region is obtained by noting the temperature $T_2$ for pressures $P>P_z$ and $T_1$ for $P<P_z$ from the Gibbs free energy plots (Fig. \ref{RPTgt}). The result is shown in \ref{RPTnbh}, where the separation line (black solid curve) has a discontinuity at $(P_z, T_z)$. In the same plot, the transition lines of zeroth-order and first-order phase transition are also given. As $T_2$ is the zeroth-order phase transition point in the pressure range $P\in (P_t,P_z)$, the corresponding transition line (solid magenta line) matches with the extension of no black hole line from $P_z$ to $P_t$. The first order coexistence line (red solid line) is also obtained from the Gibbs free energy behaviour. The first order coexistence curve and zeroth-order transition line meet at $(P_t, T_t)$, which is the \emph{triple point}. The zeroth order transition line terminates at $(P_z,T_z)$, and the other at the critical point $(P_{c1,T_{c1}})$.

\begingroup
\setlength{\tabcolsep}{8pt} 
\renewcommand{\arraystretch}{1.2} 
\begin{center}
\begin{table*}
\centering
\begin{tabular}{ cccccccc} 
\hline
\hline
$b$ & 0.43 & 0.44 & 0.45 & 0.46 & 0.47 & 0.48 & 0.49 \\
\hline
$P_{c0}$ & -0.000663 & -0.00186 & -0.003253 & -0.004888 & -0.006838 & -0.009242 & -0.012447 \\
$T_{c0}$ & 0.033189 & 0.030199 & 0.026885 & 0.023198 & 0.019061 & 0.014328 & 0.008676 \\ 
\hline
$P_{c1}$ & 0.003711 & 0.003686 & 0.003663 & 0.003643 & 0.003625 & 0.003609 & 0.003594 \\
$T_{c1}$ & 0.045414 & 0.045289 & 0.045176 & 0.045074 & 0.044981 & 0.044895 & 0.044817 \\
\hline
$P_{t}$ & 0.002225 & 0.001922 & 0.001612 & 0.001294 & 0.000971 & 0.000644 & 0.000318 \\
$T_{t}$ & 0.037435 & 0.035249 & 0.032722 & 0.029762 & 0.026208 & 0.021759 & 0.015646 \\
\hline
$P_z$ & 0.002395 & 0.002101 & 0.00179 & 0.001464 & 0.001122 & 0.000747 &  0.000357  \\
$T_z$ & 0.037658 & 0.035457 & 0.032906 & 0.029914 & 0.026322 & 0.021821 & 0.015663 \\
\hline
\end{tabular}
\caption{\label{rptcritical} The values of  $(P_{c0}, T_{c0})$, $(P_{c1}, T_{c1})$, $(P_{t}, T_{t})$ and $(P_{z}, T_{z})$ for different $b$ values, with $Q=1$, corresponding to the RPT of the black hole.}
\end{table*}
\end{center}
\endgroup

Here too we obtain the spinodal curves using the definition Eq. \ref{spinodaleqn}, which are shown in Fig. \ref{RPTPT} along with the transition lines. The spinodal curve, the extremal points of the isothermal and isobaric curves by definition, have two turning points in the reentrant case, which are marked as $c_0$ and $c_1$. As mentioned earlier, in case 3, which we have chosen, the point $c_0$ lies in the negative pressure region. The region between $(P_t,T_t)$ and $(P_z,T_z)$ is enlarged in Fig. \ref{RPTPTsub}. Note that, for  $\{P<P_z, T<T_z\} $  the spinodal curve and no black hole line are the same. The regions separated by first-order coexistence line and zeroth-order transition line are clearly seen here. As both zeroth-order and first-order transitions are between IBH and LBH both lines separate these two phases in their respective domains. In Fig. \ref{RPTTV}, the phase structure for the reentrant case is given in $T_r-V_r$ plane, where the coexistence region of IBH and LBH is clearly shown. The IBH region is on the left, and the LBH is on the right side of the coexistence curve. As in $P_r-T_r$ plane here also the spinodal curve has two turning points, at the critical points $c_0$ and $c_1$. In contrast to the case of SPT, here, the spinodal curve does not intersect the $V_r$ axis near the origin, instead shoots to infinity. On the other side, near the LBH phase, there is no change in the behaviour. The region near the origin is enlarged in the inset, where it can be seen that the IBH branch terminates when it meets the spinodal curve. Before concluding this section we list the values of $(P_{c0}, T_{c0})$, $(P_{c1}, T_{c1})$, $(P_{t}, T_{t})$ and $(P_{z}, T_{z})$ for different $b$ values for the reentrant case in table \ref{rptcritical}.


\section{Ruppeiner Geometry and Microstructure of the Black Hole}
\label{sectwo}
In this section, we investigate the microstructure of four-dimensional Born-Infeld AdS black hole using Ruppeiner geometry, constructed in the parameter space with fluctuation coordinates as temperature and volume. Particularly, we are interested in the microstructure that leads to the reentrant phase transition. As we will see, the underlying microstructure for the SPT has similarity with the RN-AdS black hole phase transition, whereas the RPT emerges due to a different nature of the black hole microstructure. 

To set up the Ruppeiner geometry, we consider two subsystems of an isolated thermodynamic system having a total entropy $S$. The smaller subsystem is assigned with an entropy $S_B$ and the larger subsystem with entropy $S_E$. Viewing the larger subsystem as a thermal bath, we can take $S_B<<S_E\approx S$. Therefore, if the system is described by a set of independent variables $x^0$ and $x^1$,
\begin{equation}
    S(x^0,x^1)=S_B(x^0,x^1)+S_E(x^0,x^1).
\end{equation}
When the system is in thermal equilibrium, the entropy $S$ is in its local maximum $S_0$. Taylor expanding the entropy in the neighbourhood of the local maximum $x^{\mu} =x^{\mu} _0$, we obtain
\begin{align}
 S=&S_0+\left. \frac{\partial S_B}{\partial x^{\mu}} \right| _{x^{\mu} _0} \Delta x^{\mu} _B+\left. \frac{\partial S_E}{\partial x^{\mu}} \right| _{x^{\mu} _0} \Delta x^{\mu} _E \nonumber \\
& + \frac{1}{2} \left. \frac{\partial ^2 S_B}{\partial x^{\mu}\partial x^{\nu}} \right| _{x^{\mu} _0} \Delta x^{\mu} _B \Delta x^{\nu} _B \nonumber \\
&+\frac{1}{2} \left. \frac{\partial ^2 S_E}{\partial x^{\mu}\partial x^{\nu}} \right| _{x^{\mu} _0} \Delta x^{\mu} _E \Delta x^{\nu} _E+\dots
\end{align}
Since the first derivatives vanish for the equilibrium condition, we have
\begin{align}
    \Delta S=S-S_0&= \frac{1}{2} \left. \frac{\partial ^2 S_B}{\partial x^{\mu}\partial x^{\nu}} \right| _{x^{\mu} _0} \Delta x^{\mu} _B \Delta x^{\nu} _B \nonumber \\
    &+\frac{1}{2} \left. \frac{\partial ^2 S_E}{\partial x^{\mu}\partial x^{\nu}} \right| _{x^{\mu} _0} \Delta x^{\mu} _E \Delta x^{\nu} _E+\dots \nonumber \\
   & \approx  \frac{1}{2} \left. \frac{\partial ^2 S_B}{\partial x^{\mu}\partial x^{\nu}} \right| _{x^{\mu} _0} \Delta x^{\mu} _B \Delta x^{\nu} _B,
\end{align}
where we have truncated the higher order terms and the second term. The second term is negligible compared to the first term, as the entropy $S_E$ of the thermal bath is close to that of the whole system, and its second derivative with respect to the intensive variables $x^{\mu}$ are smaller than those of $S_B$.

In Ruppeiner geometry, the entropy $S$ is taken as the thermodynamical potential and its fluctuation $\Delta S$ is related to the line element $\Delta l^2$, which is the measure of the distance between two neighbouring fluctuation states of the thermodynamic system \citep{Ruppeiner95}. The Ruppeiner line element is given by,

\begin{equation}
\Delta l^2=\frac{1}{k_B}g_{\mu \nu} ^R \Delta x^{\mu} \Delta x^{\nu}, 
\end{equation}
where $k_B$ is the Boltzmann constant and the metric $g_{\mu \nu} ^R$ is,
\begin{equation}
    g_{\mu \nu} ^R=-\frac{\partial ^2 S_B}{\partial x^{\mu}\partial x^{\nu}}.
\end{equation}

\begin{figure*}[t]
\centering
\subfigure[][]{\includegraphics[scale=0.9]{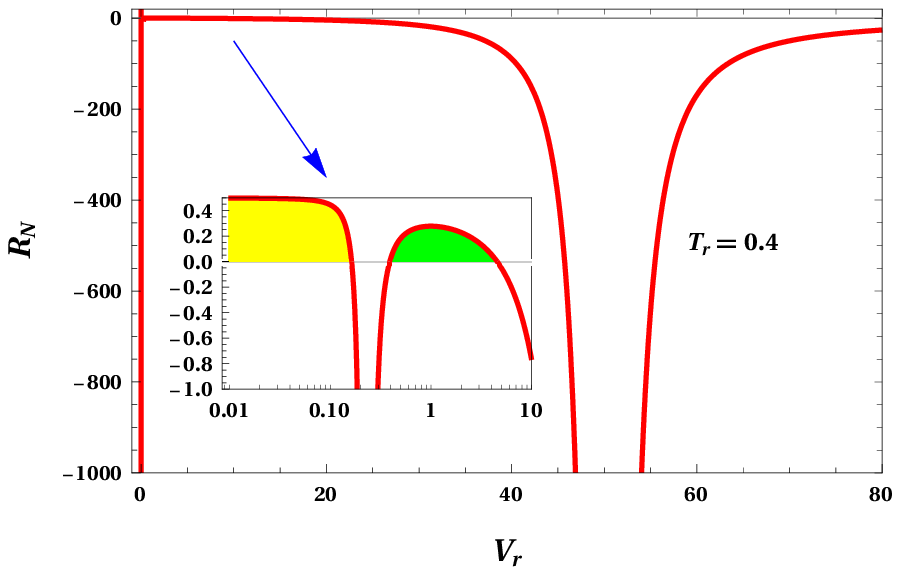}\label{SPTRV1}}
\qquad
\subfigure[][]{\includegraphics[scale=0.9]{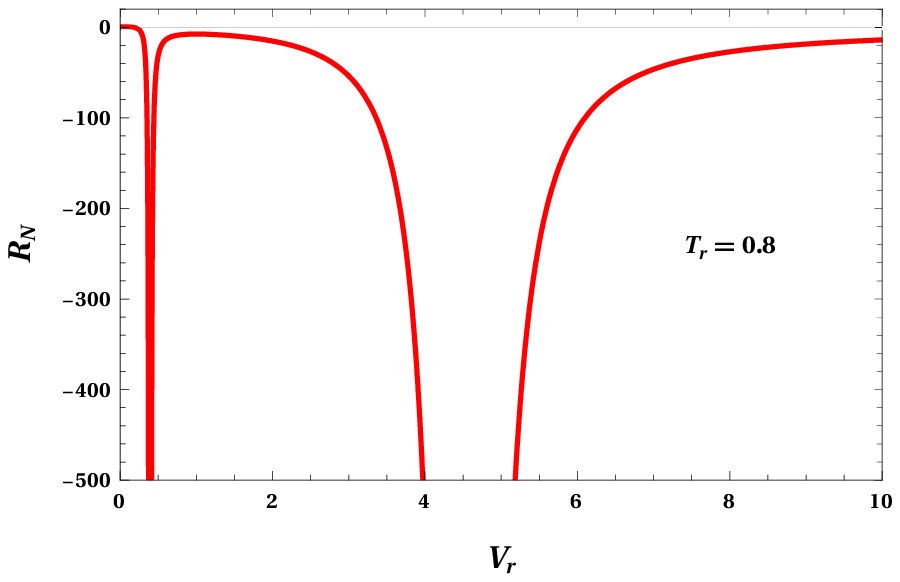}\label{SPTRV2}}
\qquad
\subfigure[][]{\includegraphics[scale=0.9]{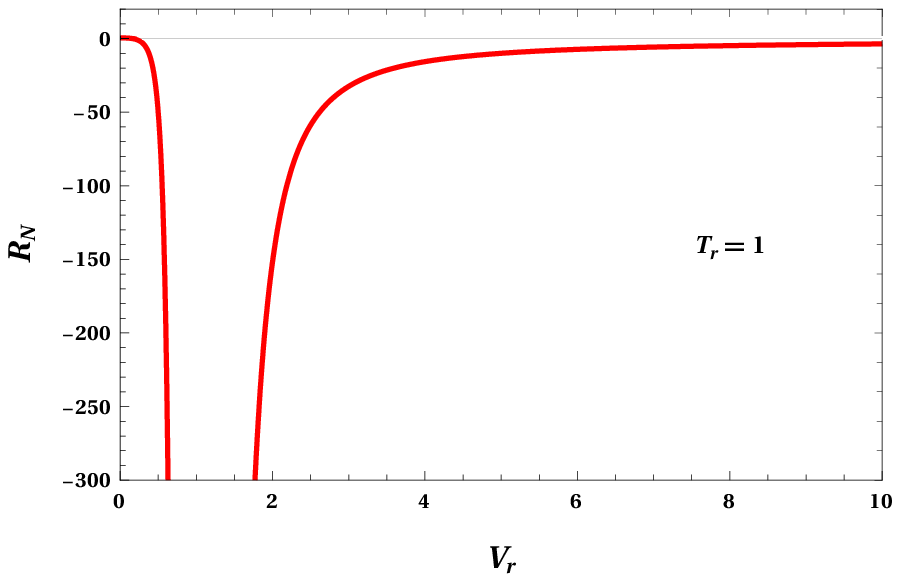}\label{SPTRV3}}
\qquad
\subfigure[][]{\includegraphics[scale=0.9]{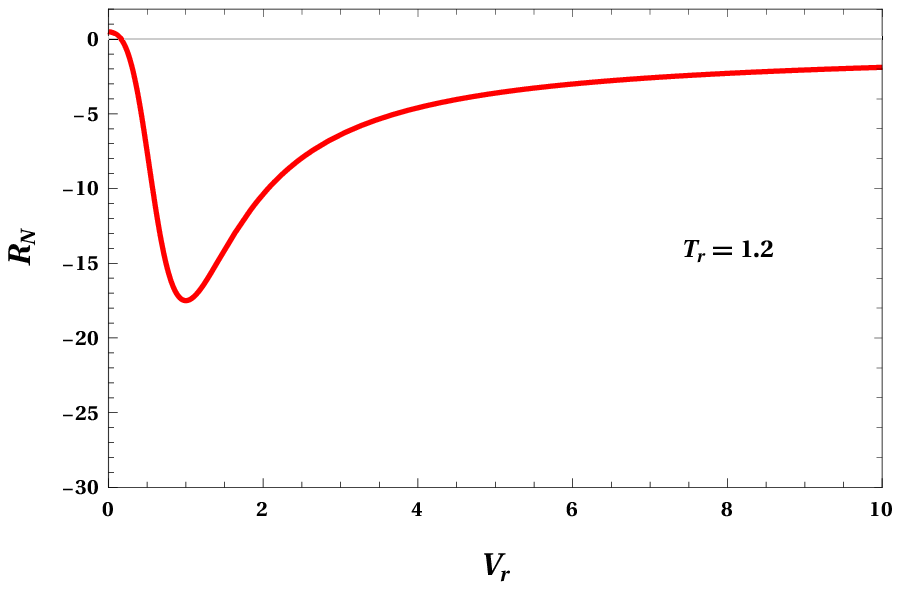}\label{SPTRV4}}

\caption{The behaviour of the normalised Ruppeiner curvature scalar $R_N$ with the reduced volume $V_r$ at a constant temperature for SPT. The insets show the enlarged portion near the origin. (Note that inset plot is in log scale). $R_N$ has positive values for small values of $V_r$, which is depicted by shaded regions. The yellow and green shaded regions correspond to SBH and coexistence phases, respectively. }
\label{SPTRN}
\end{figure*}

Since the line element is related to the distance between the neighbouring fluctuation states, the metric $g_{\mu \nu} ^R$ must be encoded with the microscopic details of the system. Taking the fluctuation coordinates $x^\mu $ as temperature and volume, and Helmholtz free energy of the system as the thermodynamic potential, the line element $\Delta l^2$ can be shown to have the following form \citep{Wei2019a},

\begin{equation}
\label{lineelement}
    \Delta l^2=\frac{C_V}{T^2}\Delta T^2-\frac{(\partial _V P)_T}{T}\Delta V^2
\end{equation}
where $C_V$ is the heat capacity at constant volume,
\begin{equation}
    C_V=T\left( \frac{\partial S}{\partial T}\right)_V.
\end{equation}

\begin{figure*}[t]
\centering
\subfigure[]{\includegraphics[scale=0.9]{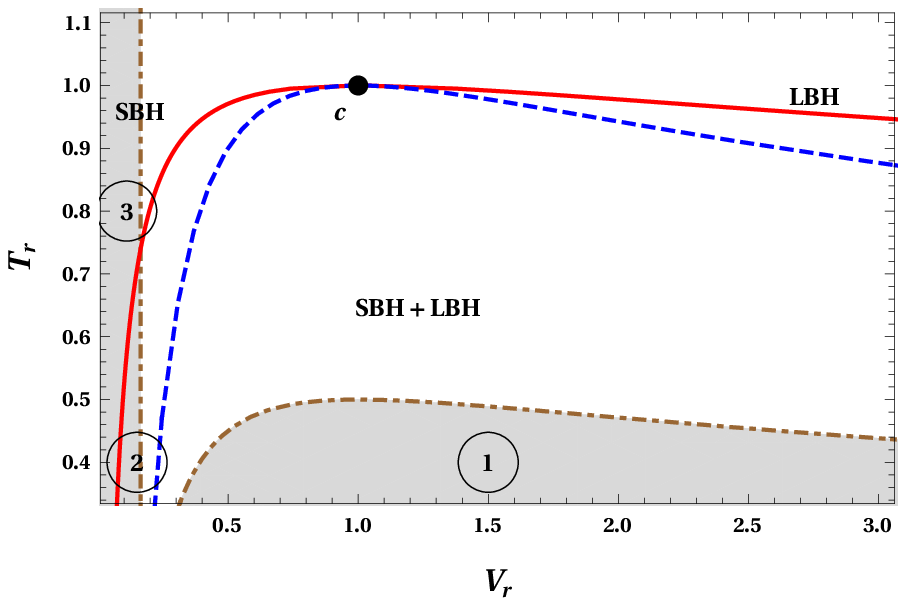}\label{SPTSign}}
\qquad
\subfigure[]{\includegraphics[scale=0.9]{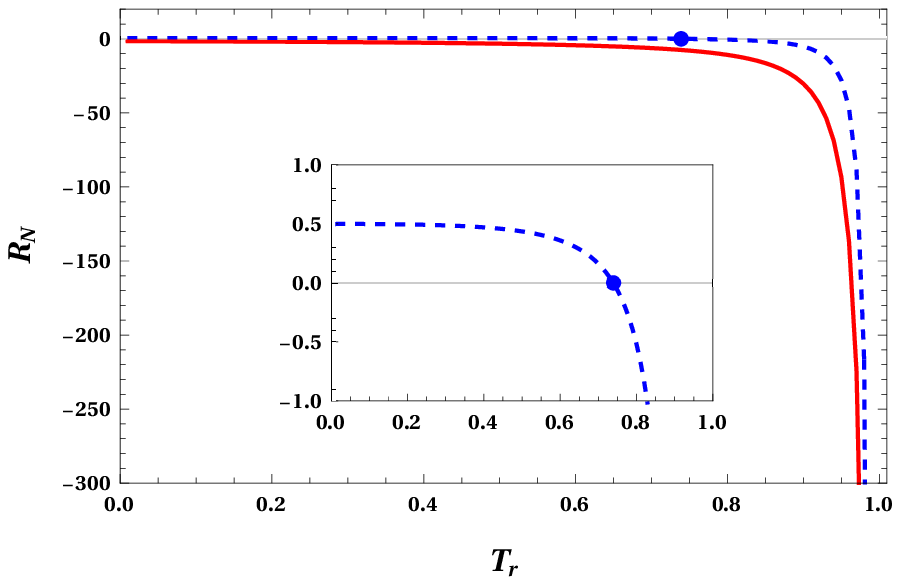}\label{SPTRT}}
\caption{ \ref{SPTSign}:  The sign changing curve of $R_N$ (brown dot-dashed line), spinodal curve (brown dashed line) and the coexistence curve (red solid line) for SPT. The shaded region (grey) corresponds to positive $R_N$, otherwise $R_N$ is negative. Region $1$ and $(2,3)$ respectively correspond to green and yellow shaded regions in Fig. \ref{SPTRV1}. The black dot at $(1,1)$ represents the critical point. \ref{SPTRT}: The behaviour of normalised Ruppeiner curvature scalar $R_N$ along the coexistence curve. The red (solid) line and  blue (dashed) line correspond to large black hole and small black hole, respectively. The change in nature of interaction of SBH is shown in inset, where $R_N$ flips its sign. Here, the blue dot corresponds to sign changing temperature.}
\end{figure*}

The Ruppeiner curvature scalar can be obtained directly from Eq. \ref{lineelement} by using the conventional definitions of Riemannian geometry. The curvature scalar so calculated is encoded with the interaction details of the microstructure of the system. The positive sign $R>0$ and negative sign $R<0$ are associated with repulsive and attractive interaction, respectively. If the Ruppeiner curvature vanishes, it implies that there is no effective interaction between the microscopic molecules. Moreover, the idea is that the Ruppeiner scalar diverges at the critical point. However, the Ruppeiner scalar obtained from the line element (\ref{lineelement}) has pathologies as the heat capacity $C_V$ vanishes for a spherically symmetric AdS black hole system. This in turn arises from the interdependence of the thermodynamic variables, namely entropy and volume. To overcome this issue, a normalised Ruppeiner scalar is defined as \citep{Wei2019a},
\begin{equation}
    R_N=C_V R
\end{equation}
where $R$ is the curvature scalar calculated from Eq. \ref{lineelement}. 
The normalised curvature scalar diverges at the critical point of phase transition.  Before proceeding further, we reiterate the following. 
\begin{itemize}
\item 
The sign of $R_N$ reveals the nature of dominant interaction in the black hole microstructure. Positive/negative for repulsive/attractive and zero for no interaction.
\item
The absolute value of $R_N$ is the measure of the average number of correlated constituents, in general. For a black hole system, this could be the average number of correlated black hole molecules.
\end{itemize}

\subsection{Standard Phase Transition Case}

\begingroup
\setlength{\tabcolsep}{8pt} 
\renewcommand{\arraystretch}{1.2} 
\begin{center}
\begin{table*}
\centering
\begin{tabular}{ cccccc } 
\hline
\hline
$b$ & 0.6 & 0.7 & 0.8 & 0.9 & 1 \\
\hline
$\alpha$ (CSSBH) & 2.06414 &  2.06982 & 2.073 & 2.07496 & 2.07629 \\
$-\beta $ (CSSBH) & 2.57849 & 2.6243 & 2.64994 & 2.66583 & 2.67658 \\ 
\hline
$\alpha$ (CSLBH) & 1.9239 & 1.9215 & 1.92011 & 1.91926 & 1.91868 \\
$-\beta $ (CSLBH) & 1.49737 & 1.47446 & 1.46131 & 1.45317 & 1.44762 \\ 
\hline
\hline
\end{tabular}
\caption{\label{sptalphabeta} The values of $\alpha$ and $\beta$ obtained by numerical fit for coexistence saturated small black holes (CSSBH) and coexistence saturated large black hole (CSLBH) for different values of $b$, with $Q=1$, for the SPT case.}
\end{table*}
\end{center}
\endgroup

\begin{figure*}[t]
\centering
\subfigure[][]{\includegraphics[scale=0.9]{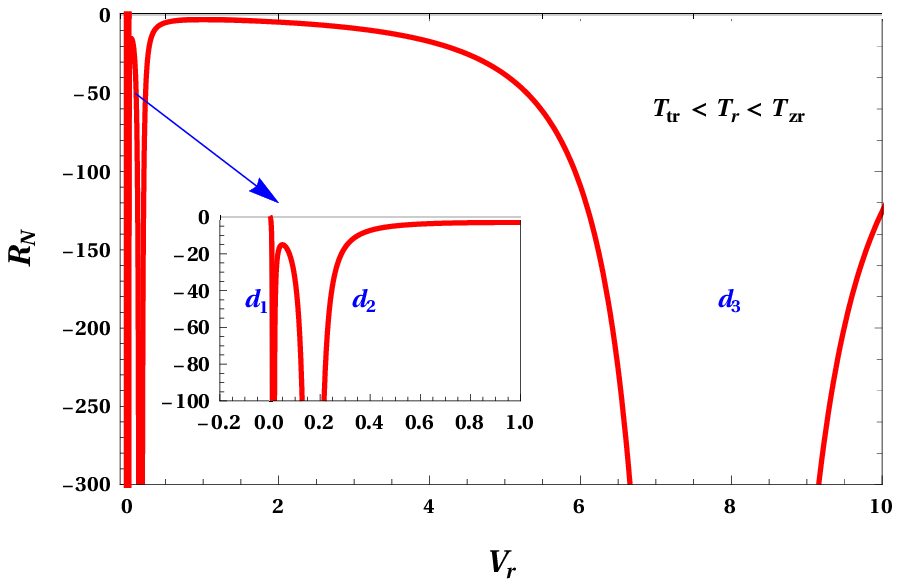}\label{RPTRV1}}
\qquad
\subfigure[][]{\includegraphics[scale=0.9]{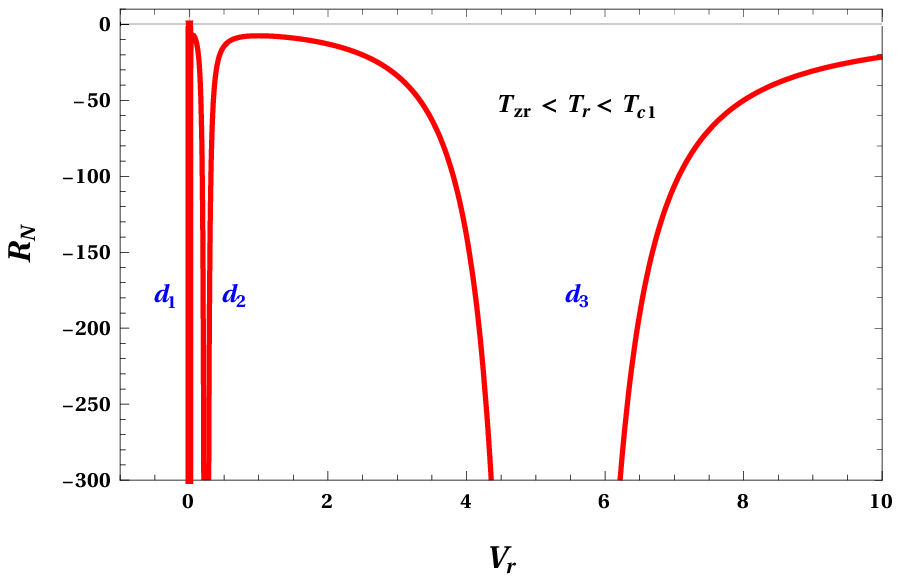}\label{RPTRV2}}
\qquad
\subfigure[][]{\includegraphics[scale=0.9]{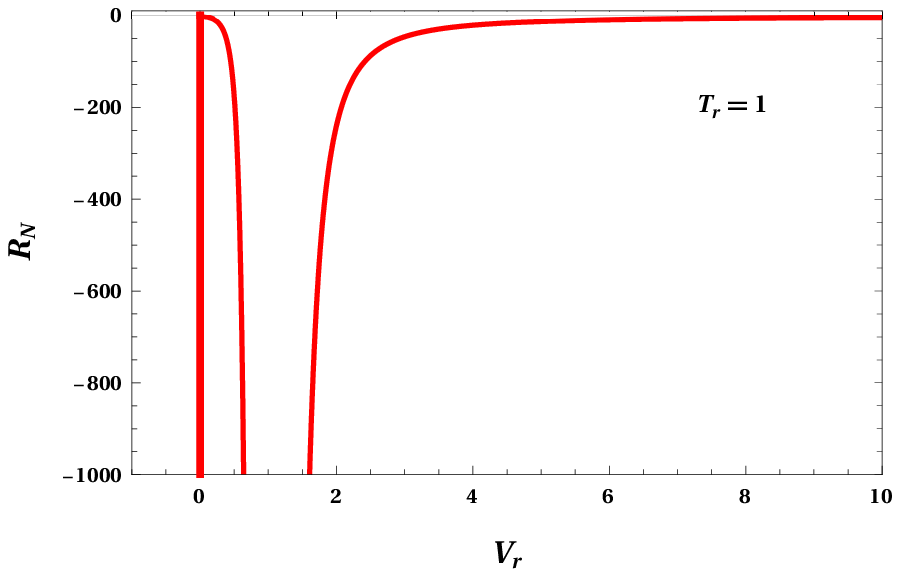}\label{RPTRV3}}
\qquad
\subfigure[][]{\includegraphics[scale=0.9]{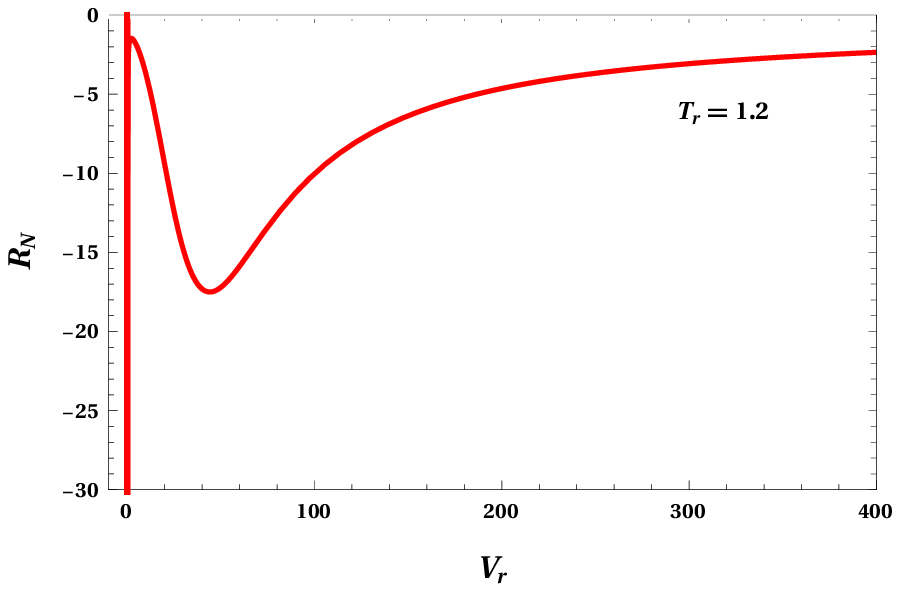}\label{RPTRV4}}
\caption{The behaviour of the normalised curvature scalar $R_N$ against the volume $V_r$ at constant temperature $T_r$ for the RPT case. For $T<T_{c1}$ there are three divergences, of which two are near the origin shown in insets of (a).  For temperature $T=T_{c1}$ one divergence disappears.  Unlike SPT case, for temperatures $T>T_{c1}$ we still have one divergence (d). Here, we take $Q=1$ and $b=0.45$.}
\label{RNRPT}
\end{figure*}

We obtained the normalised curvature scalar $R_N$ for the four-dimensional Born-Infeld AdS black hole using the above definitions. The SPT and RPT cases are analysed separately. The curvature scalar $R_N$ is expressed in terms of reduced parameters using Eq. \ref{red1} and Eq. \ref{red2} in the respective cases. First, we will analyse the SPT case. It is found that the behaviour of $R_N$ for SPT case is similar to that of RN-AdS black hole. The functional behaviour of $R_N$ against thermodynamic volume $V_r$ for a fixed temperature of $T_r$ is studied in fig. (\ref{SPTRN}). For $T_r<1$, below critical temperature, $R_N$ has two negative divergences. These two divergences gradually come closer as temperature increases and merge at $T_r=1$. The divergence at critical temperature occurs at $V_r=1$. For temperatures above the critical value, both divergences disappear. The divergences of $R_N$ are related to the spinodal curve. In fact, $R_N$ diverges along the spinodal curve. Since $V_r$ is doubly degenerate in the $T_r-V_r$ spinodal curve (Fig. \ref{SPTTV}), below critical point, it leads to two divergences for $T_r<1$. For low temperatures, we notice that $R_N$ takes positive values in a very small domain (shown in the insets of Fig. \ref{RPTRV1},  near the origin)  as $V_r\rightarrow 0$, which is the small black hole phase. We observe positive $R_N$ on both sides of the near origin divergence. The left side corresponds to the SBH phase (shaded yellow) and the right side corresponds to the coexistence phase (shaded green). The shaded yellow region exists for all temperatures $T_r$, whereas the shaded green region disappears after a particular $T_r$. This can be understood clearly using the sign-changing curve later (Fig. \ref{SPTSign}). The yellow region implies that the SBH phase has a dominant repulsive interaction. On the other hand, the LBH phase is always characterised by the dominant attractive interaction. To check whether the repulsive interaction regions are thermodynamically stable, we present the regions of $R_N$ with different sign in $T_r-V_r$ plane along with the coexistence and the spinodal curves (fig. \ref{SPTSign}). The sign-changing curve of $R_N$ is obtained by setting $R_N=0$. The solution satisfy the condition $T_{sr}=T_{spr}/2$, where $T_{0r}$ is the sign-changing temperature and $T_{spr}$ temperature along the spinodal curve. In fig. \ref{SPTSign}, the shaded region corresponds to the negative sign of $R_N$. Region $1$ is the coexistence region of SBH and LBH. Region $2$ lies in the metastable phase, which is not interesting from the thermodynamic perspective. Region $3$ includes part of SBH phase of the black hole,  implying repulsive interactions among the small black hole constituents.

\begin{figure*}[t]
\centering
\subfigure[]{\includegraphics[scale=0.9]{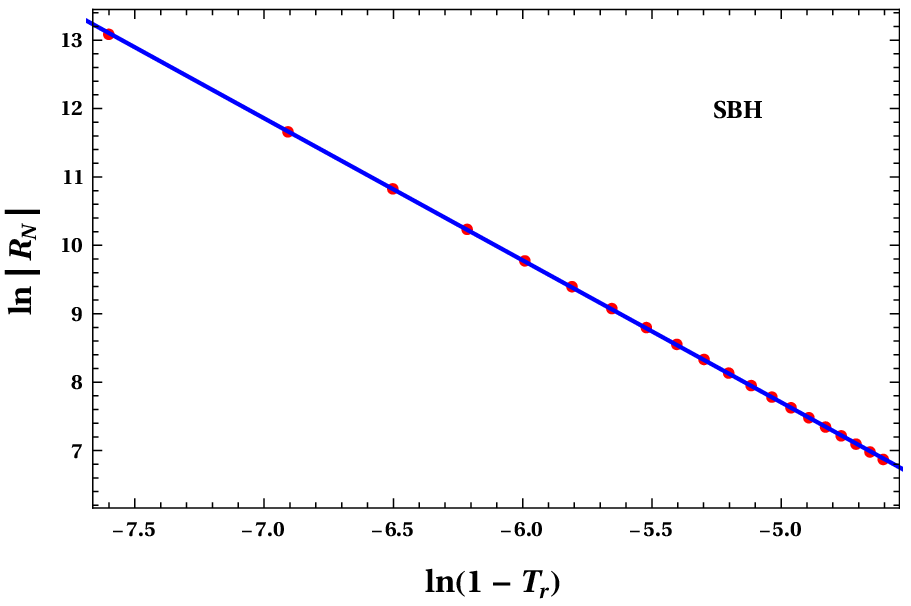}\label{SPTfit1}}
\qquad
\subfigure[]{\includegraphics[scale=0.9]{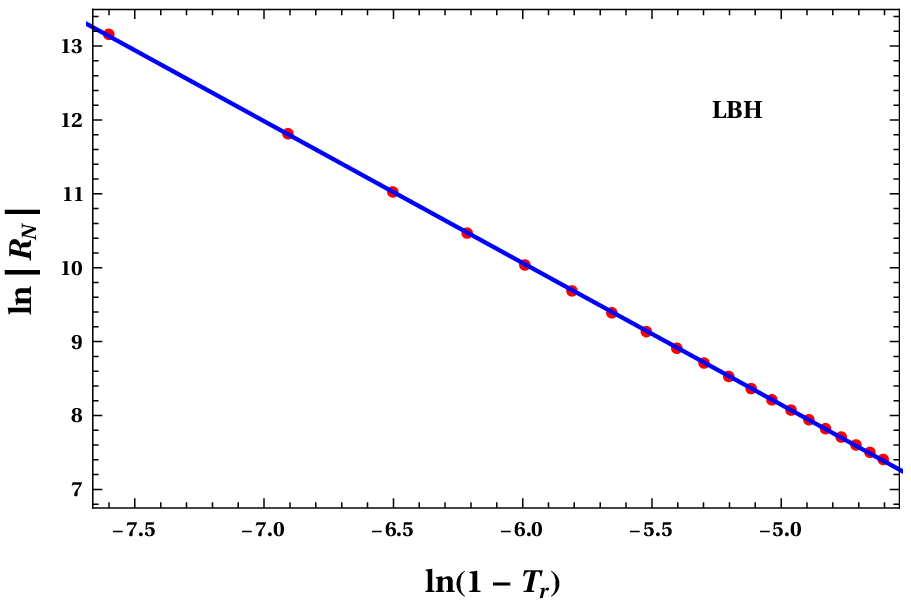}\label{SPTfit2}}
\caption{The fitting curves of $\ln R_N$ vs $\ln (1-T_r)$ near the critical point. The red dots are numerical data and blue solid lines are obtained from fitting formulas. We have varied $T_r$ from $0.99$ to $0.9999$ to obtain numerical data. (a) The coexistence saturated SBH branch (b) The coexistence saturated LBH branch}
\label{SPTfit}
\end{figure*}

We now examine the behaviour of $R_N$ along the coexistence curve and study the critical phenomena \ref{SPTRT}. The nature of interaction can be better understood from the behaviour of $R_N$ along the coexistence curve. From Fig. \ref{SPTRT} it is clear that LBH phase always has dominant attractive interaction ($R_N$ always negative), whereas, the SBH phase switches to dominant repulsive interaction ($R_N$ positive) from dominant attractive interaction ($R_N$ negative) for low temperature values. The analysis for different values of Born Infeld parameter $b$ in the SPT range was performed numerically. In the reduced coordinates, when the type of interaction in SBH switches between attractive and repulsive interaction, for $b=(0.6, 0.7, 0.8, 0.9, 1)$ the corresponding temperatures are $(0.67063, 0.708075, 0.724286, 0.733273, 0.738949)$. For temperatures above these, the absolute value of the curvature scalar of the LBH is greater than that of the SBH. This implies that the attractive interaction in the LBH is stronger than SBH. Both the SBH and LBH branches diverge to negative infinity at the critical point. This can be seen as being due to the divergence of correlation length at the critical point. This is a universal behaviour observed in other black hole systems in AdS spacetime. We obtain the critical exponent of the normalised scalar curvature $R_N$ at the critical point. Since the analytical expansion is not feasible, we assume that the scalar curvature has the following form near the critical point,
\begin{equation}
    R_N\sim (1-T_r)^{-\alpha}.
    \label{fiteqn}
\end{equation}
Taking logarithm on both sides,
\begin{equation}
   \ln |R_N|=-\alpha \ln (1-T_r)+\beta .
\end{equation}
Along the coexistence curve, for $Q=1$ and $b=1$ we obtained,
\begin{equation}
   \ln |R_N|=- 2.07629 \ln (1-T_r)-2.67658, 
\end{equation}
for the SBH branch and,
\begin{equation}
   \ln |R_N|=- 1.91868 \ln (1-T_r)-1.44762, 
\end{equation}
for the LBH branch. These are shown in Fig. \ref{SPTfit}. Considering the numerical errors we have the critical exponent $\alpha=2$. Averaging the $\beta$ values, we have,
\begin{equation}
    R_N(1-T_r)^2=e^{-(2.67658+1.44762)/2}=-0.127187\approx-\frac{1}{8}.
\end{equation}
From the numerical results given in table \ref{sptalphabeta}, we have,
\begin{align}
    R_N (1-T_r)^2 =-(&0.130298, 0.128815, 0.128012, \nonumber \\
    &0.127517, 0.127187),
\end{align}
for $b=(0.6, 0.7, 0.8, 0.9, 1)$, respectively. These values are slightly more negative than $-1/8=-0.125$. Our results are in agreement with that of both the charged AdS black hole and vdW fluid, which have a $R_N$ with a critical exponent of $2$ and constant value of $-1/8$ for $R_N (1-T_r)^2$, near the critical point, which is a universal behaviour. 

To summarise the nature of interaction in the underlying microstructure of SPT, the LBH phase is always characterised by attractive interaction (like a bosonic gas), whereas, the SBH phase can have both attractive and repulsive interaction (like  a  quantum  anyon  gas).

\subsection{Reentrant Phase Transition Case}
We analyse the underlying microstructure of the black hole that results in RPT using the same method as before. The normalised Ruppeiner curvature scalar is plotted against the volume for constant temperatures in Fig. \ref{RNRPT}. Compared to the SPT case, here we observe an additional divergence for all temperatures $T>T_{c0}$, and only one divergence for $T<T_{c0}$. This is because, the spinodal curve has a different structure in the RPT case (Fig. \ref{SPTTV}) compared to SPT case (Fig. \ref{RPTTV}). For phase transition temperatures, which lie in the range $T\in (T_t, T_{c1})$, there exist three divergences as shown in  Fig \ref{RPTRV1} and Fig. \ref{RPTRV2}. Most general conclusions on the appearance of this divergences can be learnt from the spinodal curve. Here, the left-most divergence, say $d_1$, corresponds to the spinodal curve branch that shoots to infinity after turning from $c_0$ in Fig. \ref{RPTTV}, which exists for temperatures $T>T_{c0}$. The middle divergence, say $d_2$, corresponds to the spinodal curve between $c_0$ and $c_1$, appearing in the temperature range $T\in (T_{c0},T_{c1})$. And the rightmost divergence, say $d_3$, is on the spinodal curve on the LBH side after $c_1$, which present for all temperatures $T<T_{c1}$.   At $T=T_{c1}$ divergences $d_2$ and $d_3$ merge together at $V_r=1$, leaving only in total two divergences, Fig. \ref{RPTRV3}. For completeness we mention that,  at $T_{c0}$ divergences $d_1$ and $d_2$ merge together, which is not relevant and hence not shown here. Interestingly, one divergence remains even for temperatures $T>T_{c1}$, Fig. \ref{RPTRV4}. These divergences separate different phases of the black hole from each other in the $R_N-V_r$ plane, which enables us to search for the kind of interaction in each phase. The region between $d_1$ and $d_2$ corresponds to IBH phase and right side of $d_3$ to LBH phase. Left side of $d_1$, and between $d_2$ and $d_3$ are coexistence regions. Contrary to the SPT case, $R_N$ always takes negative values, implying only a dominant attractive interaction in all of the phases in RPT case. We search for a possible repulsive interaction at low $V_r$ values, shown in the inset of Fig. \ref{RPTRV1}, which is a null result. This result is true for all phase transition temperatures. In short, for RPT case the black hole microstructure has only attractive interactions. 

\begin{figure*}[t]
\centering
\subfigure[]{\includegraphics[scale=0.9]{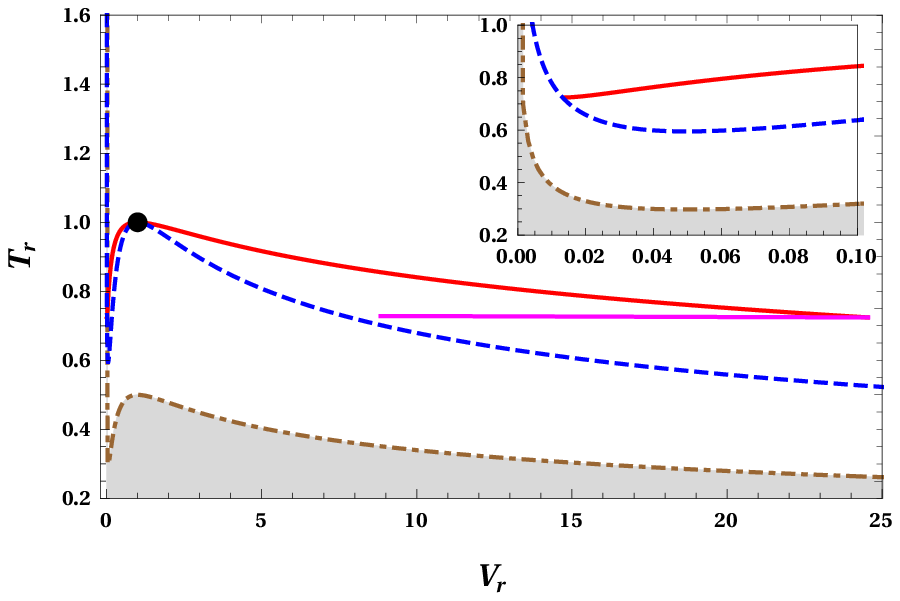}\label{RPTSign}}
\qquad
\subfigure[]{\includegraphics[scale=0.9]{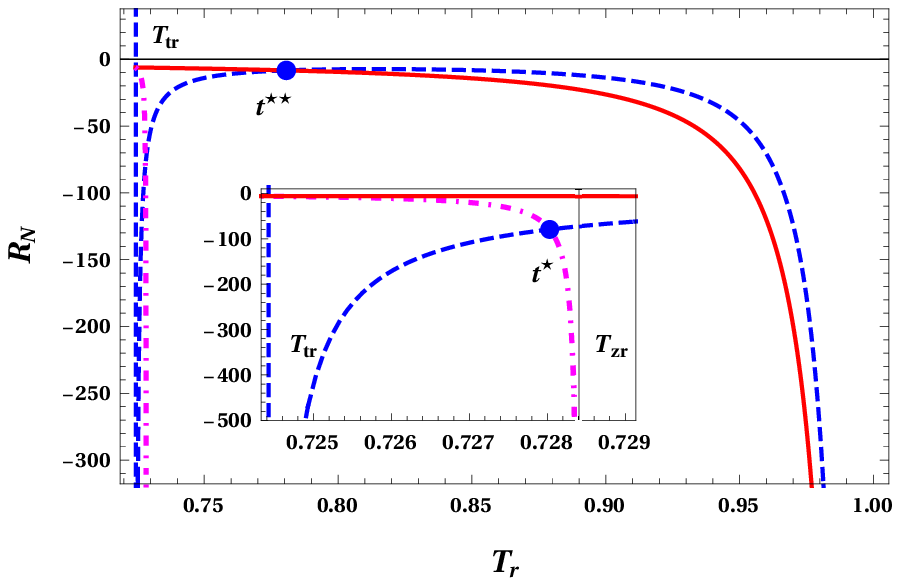}\label{RPTRT}}
\caption{ \ref{RPTSign}:  The sign changing curve (brown dot-dashed line) of $R_N$, spinodal curve (blue dashed line) and the coexistence curve (red solid line) for the RPT case. The shaded region (grey) corresponds to positive $R_N$, elsewhere $R_N$ is negative. All the stable phases, including near the zeroth order transition line (solid magenta line), lie outside the shaded region. The region near the origin is enlarged and shown in inset. \ref{RPTRT}: The behaviour of normalised curvature scalar $R_N$ along the transition line of first-order and zeroth-order phase transition.  The red (solid) line and  blue (dashed) lines correspond to LBH and IBH phases, respectively.  Both diverge at the critical point.  The IBH branch suffers a discontinuity near the triple point $T_t$. The dot-dashed magenta line between $T_{tr}$ and $T_{zr}$ (clearly shown in the inset) corresponds to the LBH phase of zeroth-order phase transition, which is a miniature version of the other LBH branch with a divergence at $T_z$, which corresponds to the termination point of the transition. In the plot, the temperatures  corresponding to the blue dots $t^\star$ and $t^{\star \star}$ are $T^{\star}_r=0.72803$ and $T^{\star \star}_r=0.780668$. We take $Q=1$ and $b=0.45$, respectively.}
\end{figure*}

\begingroup
\setlength{\tabcolsep}{8pt} 
\renewcommand{\arraystretch}{1.2} 
\begin{center}
\begin{table*}
\centering
\begin{tabular}{ ccccccccc } 
\hline
\hline
$b$ & 0.43 & 0.44 & 0.45 & 0.46 & 0.47& 0.48 &0.49\\
\hline
$\alpha$ (CSIBH) & 1.91582 & 2.01262 & 2.07636 & 1.99898 & 2.02411 & 2.03111 & 2.03868  \\
$-\beta $ (CSIBH) &  1.68789 & 2.2181 & 2.56921 & 2.17032 & 2.31204 & 2.357 & 2.40382 \\ 
\hline
$\alpha$ (CSLBH) & 2.04466 & 1.9619 & 1.91492 & 1.99174 & 1.95724 & 1.9511 & 1.94448 \\
$-\beta $ (CSLBH) & 2.19809 & 1.75127 & 1.49509 & 1.8951 & 1.70734 & 1.67 & 1.63093 \\ 
\hline
\hline
\end{tabular}
\caption{\label{rptalphabeta} The values of $\alpha$ and $\beta$ obtained by numerical fit for coexistence saturated intermediate black holes (CSIBH) and coexistence saturated large black hole (CSLBH) for different values of $b$, with $Q=1$, for the RPT case.}
\end{table*}
\end{center}
\endgroup

We confirm the above result by looking at the region covered by the sign-changing curve (Fig. \ref{RPTSign}). The positive $R_N$ values, shaded region, do not overlap with any stable state of the system. Note that the transition line of the zeroth-order phase transition (solid magenta line) also lies outside the shaded region. Near origin domain is enlarged in the inset for clarity. The behaviour of $R_N$ along the transition curve is analysed in Fig. \ref{RPTRT}. We consider the first-order phase transition coexistence curve and the zeroth-order transition curve. As the zeroth-order transition line is not a coexistence curve, it gives only one branch. We consider only the region of temperature where the phase transition takes place, $T_t<T<T_{c1}$, as below the temperature $T_t$ there is no phase transition. Once again, the positive sign of $R_N$ asserts that, both IBH and LBH phases have dominant attractive interaction (like a bosonic gas). That is, during a reentrant phase transition in BI-AdS black hole, both the zeroth-order and first-order transition preserve the nature of the interaction between the microstructures. However, we notice that the correlation between the constituents changes at different temperatures. The flipping temperatures are clearly shown in Fig. \ref{RPTRT} in blue dots. For zeroth-order transition (jumping from blue dashed line to magenta dot-dashed line and vice versa), IBH phase has higher correlation strength than LBH in the temperature range $T\in (T_t, T^\star)$. This property is reversed in the region  $T\in (T^\star,T_z)$. On the other hand, for the first-order phase transition (jumping from blue dashed line to red solid line and vice versa), IBH phase exhibits higher correlation among the constituents than LBH for temperatures  $T\in (T_t, T^{\star \star})$, which is flipped for  $T\in (T^{\star \star}, T_{c1})$. In fact, the change in the strength of the interaction is also present in SPT case. Interestingly, near the triple point, IBH phase shows a huge surge in correlation, whereas the correlation in LBH phase here, is very weak. Also, the red and magenta line merge near $T_t$, signifying the meeting and termination of zeroth-order and first-order transition line at that point. As in SPT case near the critical point the diverging correlation length leads to diverging $R_N$. For magenta line it happens at $T_z$, in a sense $(P_z,T_z)$ acts as a critical point (see for example \citep{Altamirano:2013ane}). During a RPT, the initial and final phases are the same from the macroscopic point of view. Microscopically, this is almost true for $T<T^\star$, where red and magenta line (initial and final states in RPT) are close together. However, for $T>T^\star$ there is a difference in strength of interaction, which becomes considerable near $T^\star$.

The critical phenomena of $R_N$ is studied for different values of $b$ numerically using Eq. \ref{fiteqn}. From the numerical results given in table \ref{rptalphabeta}, we have,
\begin{align}
    R_N (1-T_r)^2 =-(&0.143275, 0.137424, 0.131053, 0.13098,\nonumber\\
    &0.13403, 0.13352, 0.133004),
\end{align}
for $b=(0.43, 0.44, 0.45, 0.46, 0.47, 0.48, 0.49)$, respectively. As in the case of SPT, these values are slightly more negative than $-1/8$. Within the numerical errors we have obtained the universal constant  $R_N (1-T_r)^2$ as $-1/8$ and the critical exponent $2$ in RPT case too. To cut a long story short, we observe the existence of only homogeneous dominant attractive interaction in the black hole for RPT case, and both repulsive and attractive interaction in SPT case. This difference in the microstructure interaction for SPT and RPT in BI-AdS black hole tells that the microstructure is determined by the coupling parameter $b$. 

\section{Discussions}
\label{secthree}

In this article, we have studied the microstructure of four-dimensional Born-Infeld AdS black hole employing a novel Ruppeiner geometry method. As the entropy and volume are interdependent in a spherically symmetric AdS black hole, the application of Ruppeiner geometry for such a system must be carried out with utmost care. Keeping this in mind, we have constructed the Ruppeiner curvature scalar in the parametric space where the temperature and volume are the fluctuation coordinates. We investigated the phase structure and the corresponding microscopic interactions for both the standard van der Waals (SPT) and the reentrant phase transitions (RPT), where a first-order phase transition is accompanied by a zeroth-order transition. We found that the microstructure that leads to RPT is distinct from that of SPT. Our study shows that the Born Infeld coupling coefficient $b$ determines the microscopic interaction of the black hole. Since the analytical investigation is not possible due to the complexity of the spacetime, we have carried out the study numerically.

In the first part of the article, we investigated the phase structure of the black hole using the spinodal and transition curves in SPT and RPT cases. The black hole has four different cases depending on the value of $b$, and shows the distinct RPT for certain pressure range. There are two RPT cases with two critical points in each. It is found that the phase structure associated with SPT and RPT are distinct. The phase diagrams are presented in pressure-temperature ($P-T$) and temperature-volume ($T-V$) planes, where, stable, metastable and coexistence phases are studied. In both SPT and RPT cases, we conveniently define the reduced parameters and all analyses were carried out in terms of them.

The second half of the article is devoted to the study of microstructure, which reveal distinct microstructure for SPT and RPT cases. Two important features, the nature of interaction and strength of correlation are sought for in this study. The microstructure that corresponds to SPT is analogous to that of RN-AdS black hole. The small black hole phase shows a dominant repulsive interaction in a certain range of parameters. As the interaction in this phase is attractive in other domains of parameter space, the interaction type resembles that of a quantum anyon gas. The large black hole phase is always characterised by the dominant attractive interaction, like a bosonic gas. These are inferred from the sign of the Ruppeiner curvature scalar $R_N$. For both small black hole and large black hole branches, it diverges near the critical point.  The RPT case has a different microstructure compared to the SPT case, wherein no repulsive interaction is present for both the intermediate and large black hole phases. The nature of microstructure does not change during the zeroth-order and first-order phase transition in RPT, which as a dominant attractive interaction. This suggests that the nature of interaction in RPT case is always like that of a bosonic gas. Both, intermediate and large black hole, branches diverge near the physically meaningful critical point. The signature of RPT is reflected in the behaviour of curvature scalar, through its anomalies near the termination points of the zeroth-order phase transition. The strength of interaction depends on temperature, which is inferred from the magnitude of the curvature scalar, the high correlation phases are flipped at certain temperatures for both zeroth-order and first-order transition. In fact, this reversal of correlation strength is true for SPT case too. The critical phenomenon of the system is investigated via curvature scalar. The universal properties, critical exponent-$2$ and the constant $R_N(1-T_r)^2=-1/8$, are obtained for SPT and RPT cases over a range of $b$ values. We believe that this study will help us shed more light on the black hole microstructure in general.


\acknowledgments
Authors N.K.A., A.R.C.L. and K.H. would like to thank U.G.C. Govt. of India for financial assistance under UGC-NET-SRF scheme.  M.S.A.'s research is supported by the ISIRD grant 9-252/2016/IITRPR/708.


  \bibliography{BibTex}

\providecommand{\href}[2]{#2}\begingroup\raggedright\begin{thebibliography}{10}

\bibitem{Hawking:1974sw}
S.~W. Hawking, \emph{{Particle Creation by Black Holes}},
  \href{https://doi.org/10.1007/BF02345020, 10.1007/BF01608497}{\emph{Commun.
  Math. Phys.} {\bfseries 43} (1975) 199}.

\bibitem{Bekenstein1972}
J.~D. Bekenstein, \emph{{Black holes and the second law}},
  \href{https://doi.org/10.1007/BF02757029}{\emph{Lett. Nuovo Cim.} {\bfseries
  4} (1972) 737}.

\bibitem{Bekenstein1973}
J.~D. Bekenstein, \emph{{Black holes and entropy}},
  \href{https://doi.org/10.1103/PhysRevD.7.2333}{\emph{Phys. Rev.} {\bfseries
  D7} (1973) 2333}.

\bibitem{Bardeen1973}
J.~M. Bardeen, B.~Carter and S.~W. Hawking, \emph{{The Four laws of black hole
  mechanics}}, \href{https://doi.org/10.1007/BF01645742}{\emph{Commun. Math.
  Phys.} {\bfseries 31} (1973) 161}.

\bibitem{Kastor:2009wy}
D.~Kastor, S.~Ray and J.~Traschen, \emph{{Enthalpy and the Mechanics of AdS
  Black Holes}},
  \href{https://doi.org/10.1088/0264-9381/26/19/195011}{\emph{Class. Quant.
  Grav.} {\bfseries 26} (2009) 195011}
  [\href{https://arxiv.org/abs/0904.2765}{{\ttfamily 0904.2765}}].

\bibitem{Dolan:2011xt}
B.~P. Dolan, \emph{{Pressure and volume in the first law of black hole
  thermodynamics}},
  \href{https://doi.org/10.1088/0264-9381/28/23/235017}{\emph{Class. Quant.
  Grav.} {\bfseries 28} (2011) 235017}
  [\href{https://arxiv.org/abs/1106.6260}{{\ttfamily 1106.6260}}].

\bibitem{Kubiznak2012}
D.~Kubiznak and R.~B. Mann, \emph{{P-V criticality of charged AdS black
  holes}}, \href{https://doi.org/10.1007/JHEP07(2012)033}{\emph{JHEP}
  {\bfseries 07} (2012) 033} [\href{https://arxiv.org/abs/1205.0559}{{\ttfamily
  1205.0559}}].

\bibitem{Gunasekaran2012}
S.~Gunasekaran, R.~B. Mann and D.~Kubiznak, \emph{{Extended phase space
  thermodynamics for charged and rotating black holes and Born-Infeld vacuum
  polarization}}, \href{https://doi.org/10.1007/JHEP11(2012)110}{\emph{JHEP}
  {\bfseries 11} (2012) 110} [\href{https://arxiv.org/abs/1208.6251}{{\ttfamily
  1208.6251}}].

\bibitem{Kubiznak:2016qmn}
D.~Kubiznak, R.~B. Mann and M.~Teo, \emph{{Black hole chemistry: thermodynamics
  with Lambda}}, \href{https://doi.org/10.1088/1361-6382/aa5c69}{\emph{Class.
  Quant. Grav.} {\bfseries 34} (2017) 063001}
  [\href{https://arxiv.org/abs/1608.06147}{{\ttfamily 1608.06147}}].

\bibitem{hudson1904gegenseitige}
C.~S. Hudson, \emph{Die gegenseitige l{\"o}slichkeit von nikotin in wasser},
  {\emph{Zeitschrift f{\"u}r Physikalische Chemie} {\bfseries 47} (1904) 113}.

\bibitem{1994PhR249135N}
T.~{Narayanan} and A.~{Kumar}, \emph{{Reentrant phase transitions in
  multicomponent liquid mixtures}},
  \href{https://doi.org/10.1016/0370-1573(94)90015-9}{\emph{physrep} {\bfseries
  249} (1994) 135}.

\bibitem{2016JPhCS.670a2040P}
O.~{Panella} and P.~{Roy}, \emph{{Re-entrant phase transitions in
  non-commutative quantum mechanics}},  in \emph{Journal of Physics Conference
  Series}, vol.~670 of \emph{Journal of Physics Conference Series}, p.~012040,
  Jan., 2016, \href{https://doi.org/10.1088/1742-6596/670/1/012040}{DOI}.

\bibitem{Zou:2013owa}
D.-C. Zou, S.-J. Zhang and B.~Wang, \emph{{Critical behavior of Born-Infeld AdS
  black holes in the extended phase space thermodynamics}},
  \href{https://doi.org/10.1103/PhysRevD.89.044002}{\emph{Phys. Rev. D}
  {\bfseries 89} (2014) 044002}
  [\href{https://arxiv.org/abs/1311.7299}{{\ttfamily 1311.7299}}].

\bibitem{Altamirano:2013ane}
N.~Altamirano, D.~Kubiznak and R.~B. Mann, \emph{{Reentrant phase transitions
  in rotating anti--de Sitter black holes}},
  \href{https://doi.org/10.1103/PhysRevD.88.101502}{\emph{Phys. Rev. D}
  {\bfseries 88} (2013) 101502}
  [\href{https://arxiv.org/abs/1306.5756}{{\ttfamily 1306.5756}}].

\bibitem{Altamirano:2013uqa}
N.~Altamirano, D.~Kubiz\v~nák, R.~B. Mann and Z.~Sherkatghanad,
  \emph{{Kerr-AdS analogue of triple point and solid/liquid/gas phase
  transition}},
  \href{https://doi.org/10.1088/0264-9381/31/4/042001}{\emph{Class. Quant.
  Grav.} {\bfseries 31} (2014) 042001}
  [\href{https://arxiv.org/abs/1308.2672}{{\ttfamily 1308.2672}}].

\bibitem{Altamirano:2014tva}
N.~Altamirano, D.~Kubiznak, R.~B. Mann and Z.~Sherkatghanad,
  \emph{{Thermodynamics of rotating black holes and black rings: phase
  transitions and thermodynamic volume}},
  \href{https://doi.org/10.3390/galaxies2010089}{\emph{Galaxies} {\bfseries 2}
  (2014) 89} [\href{https://arxiv.org/abs/1401.2586}{{\ttfamily 1401.2586}}].

\bibitem{Kubiznak:2015bya}
D.~Kubiznak and F.~Simovic, \emph{{Thermodynamics of horizons: de Sitter black
  holes and reentrant phase transitions}},
  \href{https://doi.org/10.1088/0264-9381/33/24/245001}{\emph{Class. Quant.
  Grav.} {\bfseries 33} (2016) 245001}
  [\href{https://arxiv.org/abs/1507.08630}{{\ttfamily 1507.08630}}].

\bibitem{Frassino:2014pha}
A.~M. Frassino, D.~Kubiznak, R.~B. Mann and F.~Simovic, \emph{{Multiple
  Reentrant Phase Transitions and Triple Points in Lovelock Thermodynamics}},
  \href{https://doi.org/10.1007/JHEP09(2014)080}{\emph{JHEP} {\bfseries 09}
  (2014) 080} [\href{https://arxiv.org/abs/1406.7015}{{\ttfamily 1406.7015}}].

\bibitem{Wei:2014hba}
S.-W. Wei and Y.-X. Liu, \emph{{Triple points and phase diagrams in the
  extended phase space of charged Gauss-Bonnet black holes in AdS space}},
  \href{https://doi.org/10.1103/PhysRevD.90.044057}{\emph{Phys. Rev. D}
  {\bfseries 90} (2014) 044057}
  [\href{https://arxiv.org/abs/1402.2837}{{\ttfamily 1402.2837}}].

\bibitem{Hennigar:2015esa}
R.~A. Hennigar, W.~G. Brenna and R.~B. Mann, \emph{{$P - v$ criticality in
  quasitopological gravity}},
  \href{https://doi.org/10.1007/JHEP07(2015)077}{\emph{JHEP} {\bfseries 07}
  (2015) 077} [\href{https://arxiv.org/abs/1505.05517}{{\ttfamily
  1505.05517}}].

\bibitem{Sherkatghanad:2014hda}
Z.~Sherkatghanad, B.~Mirza, Z.~Mirzaiyan and S.~A. Hosseini~Mansoori,
  \emph{{Critical behaviors and phase transitions of black holes in higher
  order gravities and extended phase spaces}},
  \href{https://doi.org/10.1142/S0218271817500171}{\emph{Int. J. Mod. Phys. D}
  {\bfseries 26} (2016) 1750017}
  [\href{https://arxiv.org/abs/1412.5028}{{\ttfamily 1412.5028}}].

\bibitem{Hennigar:2015wxa}
R.~A. Hennigar and R.~B. Mann, \emph{{Reentrant phase transitions and van der
  Waals behaviour for hairy black holes}},
  \href{https://doi.org/10.3390/e17127862}{\emph{Entropy} {\bfseries 17} (2015)
  8056} [\href{https://arxiv.org/abs/1509.06798}{{\ttfamily 1509.06798}}].

\bibitem{Dehyadegari:2017hvd}
A.~Dehyadegari and A.~Sheykhi, \emph{{Reentrant phase transition of
  Born-Infeld-AdS black holes}},
  \href{https://doi.org/10.1103/PhysRevD.98.024011}{\emph{Phys. Rev. D}
  {\bfseries 98} (2018) 024011}
  [\href{https://arxiv.org/abs/1711.01151}{{\ttfamily 1711.01151}}].

\bibitem{Xu:2019yub}
Y.-M. Xu, H.-M. Wang, Y.-X. Liu and S.-W. Wei, \emph{{Photon sphere and
  reentrant phase transition of charged Born-Infeld-AdS black holes}},
  \href{https://doi.org/10.1103/PhysRevD.100.104044}{\emph{Phys. Rev. D}
  {\bfseries 100} (2019) 104044}
  [\href{https://arxiv.org/abs/1906.03334}{{\ttfamily 1906.03334}}].

\bibitem{Weinhold75}
F.~Weinhold, \emph{Metric geometry of equilibrium thermodynamics},
  \href{https://doi.org/10.1063/1.431689}{\emph{The Journal of Chemical
  Physics} {\bfseries 63} (1975) 2479}
  [\href{https://arxiv.org/abs/https://doi.org/10.1063/1.431689}{{\ttfamily
  https://doi.org/10.1063/1.431689}}].

\bibitem{Ruppeiner95}
G.~Ruppeiner, \emph{{Riemannian geometry in thermodynamic fluctuation theory}},
  \href{https://doi.org/10.1103/RevModPhys.67.605}{\emph{Rev. Mod. Phys.}
  {\bfseries 67} (1995) 605}.

\bibitem{Ruppeiner79}
G.~Ruppeiner, \emph{Thermodynamics: A riemannian geometric model},
  \href{https://doi.org/10.1103/PhysRevA.20.1608}{\emph{Phys. Rev. A}
  {\bfseries 20} (1979) 1608}.

\bibitem{Janyszek1990b}
H.~{Janyszek}, \emph{{Riemannian geometry and stability of thermodynamical
  equilibrium systems}},
  \href{https://doi.org/10.1088/0305-4470/23/4/017}{\emph{Journal of Physics A
  Mathematical General} {\bfseries 23} (1990) 477}.

\bibitem{Ruppeiner81}
G.~Ruppeiner, \emph{Application of riemannian geometry to the thermodynamics of
  a simple fluctuating magnetic system},
  \href{https://doi.org/10.1103/PhysRevA.24.488}{\emph{Phys. Rev. A} {\bfseries
  24} (1981) 488}.

\bibitem{Janyszek89}
H.~Janyszek and R.~Mrugal/a, \emph{Riemannian geometry and the thermodynamics
  of model magnetic systems},
  \href{https://doi.org/10.1103/PhysRevA.39.6515}{\emph{Phys. Rev. A}
  {\bfseries 39} (1989) 6515}.

\bibitem{Janyszek_1990}
H.~Janyszek and R.~Mrugaa, \emph{Riemannian geometry and stability of ideal
  quantum gases},
  \href{https://doi.org/10.1088/0305-4470/23/4/016}{\emph{Journal of Physics A:
  Mathematical and General} {\bfseries 23} (1990) 467}.

\bibitem{Oshima_1999x}
H.~Oshima, T.~Obata and H.~Hara, \emph{Riemann scalar curvature of ideal
  quantum gases obeying gentile's statistics},
  \href{https://doi.org/10.1088/0305-4470/32/36/302}{\emph{Journal of Physics
  A: Mathematical and General} {\bfseries 32} (1999) 6373}.

\bibitem{Mirza2008}
B.~Mirza and H.~Mohammadzadeh, \emph{Ruppeiner geometry of anyon gas},
  \href{https://doi.org/10.1103/PhysRevE.78.021127}{\emph{Phys. Rev. E}
  {\bfseries 78} (2008) 021127}.

\bibitem{May2013}
H.-O. May, P.~Mausbach and G.~Ruppeiner, \emph{Thermodynamic curvature for
  attractive and repulsive intermolecular forces},
  \href{https://doi.org/10.1103/PhysRevE.88.032123}{\emph{Phys. Rev. E}
  {\bfseries 88} (2013) 032123}.

\bibitem{Ruppeiner:2013yca}
G.~Ruppeiner, \emph{{Thermodynamic curvature and black holes}},
  \href{https://doi.org/10.1007/978-3-319-03774-5\_10}{\emph{Springer Proc.
  Phys.} {\bfseries 153} (2014) 179}
  [\href{https://arxiv.org/abs/1309.0901}{{\ttfamily 1309.0901}}].

\bibitem{Cai:1998ep}
R.-G. Cai and J.-H. Cho, \emph{{Thermodynamic curvature of the BTZ black
  hole}}, \href{https://doi.org/10.1103/PhysRevD.60.067502}{\emph{Phys. Rev. D}
  {\bfseries 60} (1999) 067502}
  [\href{https://arxiv.org/abs/hep-th/9803261}{{\ttfamily hep-th/9803261}}].

\bibitem{Aman:2003ug}
J.~E. Aman, I.~Bengtsson and N.~Pidokrajt, \emph{{Geometry of black hole
  thermodynamics}}, \href{https://doi.org/10.1023/A:1026058111582}{\emph{Gen.
  Rel. Grav.} {\bfseries 35} (2003) 1733}
  [\href{https://arxiv.org/abs/gr-qc/0304015}{{\ttfamily gr-qc/0304015}}].

\bibitem{Mirza:2007ev}
B.~Mirza and M.~Zamani-Nasab, \emph{{Ruppeiner Geometry of RN Black Holes: Flat
  or Curved?}},
  \href{https://doi.org/10.1088/1126-6708/2007/06/059}{\emph{JHEP} {\bfseries
  06} (2007) 059} [\href{https://arxiv.org/abs/0706.3450}{{\ttfamily
  0706.3450}}].

\bibitem{Sahay:2010wi}
A.~Sahay, T.~Sarkar and G.~Sengupta, \emph{{Thermodynamic Geometry and Phase
  Transitions in Kerr-Newman-AdS Black Holes}},
  \href{https://doi.org/10.1007/JHEP04(2010)118}{\emph{JHEP} {\bfseries 04}
  (2010) 118} [\href{https://arxiv.org/abs/1002.2538}{{\ttfamily 1002.2538}}].

\bibitem{Chaturvedi:2014vpa}
P.~Chaturvedi, A.~Das and G.~Sengupta, \emph{{Thermodynamic Geometry and Phase
  Transitions of Dyonic Charged AdS Black Holes}},
  \href{https://doi.org/10.1140/epjc/s10052-017-4678-z}{\emph{Eur. Phys. J. C}
  {\bfseries 77} (2017) 110} [\href{https://arxiv.org/abs/1412.3880}{{\ttfamily
  1412.3880}}].

\bibitem{Wei:2017icx}
S.-W. Wei, B.~Liang and Y.-X. Liu, \emph{{Critical phenomena and chemical
  potential of a charged AdS black hole}},
  \href{https://doi.org/10.1103/PhysRevD.96.124018}{\emph{Phys. Rev. D}
  {\bfseries 96} (2017) 124018}
  [\href{https://arxiv.org/abs/1705.08596}{{\ttfamily 1705.08596}}].

\bibitem{Chaturvedi:2017vgq}
P.~Chaturvedi, S.~Mondal and G.~Sengupta, \emph{{Thermodynamic Geometry of
  Black Holes in the Canonical Ensemble}},
  \href{https://doi.org/10.1103/PhysRevD.98.086016}{\emph{Phys. Rev. D}
  {\bfseries 98} (2018) 086016}
  [\href{https://arxiv.org/abs/1705.05002}{{\ttfamily 1705.05002}}].

\bibitem{Quevedo:2006xk}
H.~Quevedo, \emph{{Geometrothermodynamics}},
  \href{https://doi.org/10.1063/1.2409524}{\emph{J. Math. Phys.} {\bfseries 48}
  (2007) 013506} [\href{https://arxiv.org/abs/physics/0604164}{{\ttfamily
  physics/0604164}}].

\bibitem{Liu:2010sz}
H.~Liu, H.~Lu, M.~Luo and K.-N. Shao, \emph{{Thermodynamical Metrics and Black
  Hole Phase Transitions}},
  \href{https://doi.org/10.1007/JHEP12(2010)054}{\emph{JHEP} {\bfseries 12}
  (2010) 054} [\href{https://arxiv.org/abs/1008.4482}{{\ttfamily 1008.4482}}].

\bibitem{Niu:2011tb}
C.~Niu, Y.~Tian and X.-N. Wu, \emph{{Critical Phenomena and Thermodynamic
  Geometry of RN-AdS Black Holes}},
  \href{https://doi.org/10.1103/PhysRevD.85.024017}{\emph{Phys. Rev. D}
  {\bfseries 85} (2012) 024017}
  [\href{https://arxiv.org/abs/1104.3066}{{\ttfamily 1104.3066}}].

\bibitem{Wei:2012ui}
S.-W. Wei and Y.-X. Liu, \emph{{Critical phenomena and thermodynamic geometry
  of charged Gauss-Bonnet AdS black holes}},
  \href{https://doi.org/10.1103/PhysRevD.87.044014}{\emph{Phys. Rev. D}
  {\bfseries 87} (2013) 044014}
  [\href{https://arxiv.org/abs/1209.1707}{{\ttfamily 1209.1707}}].

\bibitem{Banerjee:2011cz}
R.~Banerjee and D.~Roychowdhury, \emph{{Critical phenomena in Born-Infeld AdS
  black holes}}, \href{https://doi.org/10.1103/PhysRevD.85.044040}{\emph{Phys.
  Rev. D} {\bfseries 85} (2012) 044040}
  [\href{https://arxiv.org/abs/1111.0147}{{\ttfamily 1111.0147}}].

\bibitem{Mansoori:2013pna}
S.~A.~H. Mansoori and B.~Mirza, \emph{{Correspondence of phase transition
  points and singularities of thermodynamic geometry of black holes}},
  \href{https://doi.org/10.1140/epjc/s10052-013-2681-6}{\emph{Eur. Phys. J. C}
  {\bfseries 74} (2014) 2681}
  [\href{https://arxiv.org/abs/1308.1543}{{\ttfamily 1308.1543}}].

\bibitem{Mo:2013sxa}
J.-X. Mo, X.-X. Zeng, G.-Q. Li, X.~Jiang and W.-B. Liu, \emph{{A unified phase
  transition picture of the charged topological black hole in Horava-Lifshitz
  gravity}}, \href{https://doi.org/10.1007/JHEP10(2013)056}{\emph{JHEP}
  {\bfseries 10} (2013) 056} [\href{https://arxiv.org/abs/1404.2497}{{\ttfamily
  1404.2497}}].

\bibitem{Mansoori:2014oia}
S.~A.~H. Mansoori, B.~Mirza and M.~Fazel, \emph{{Hessian matrix, specific
  heats, Nambu brackets, and thermodynamic geometry}},
  \href{https://doi.org/10.1007/JHEP04(2015)115}{\emph{JHEP} {\bfseries 04}
  (2015) 115} [\href{https://arxiv.org/abs/1411.2582}{{\ttfamily 1411.2582}}].

\bibitem{Hendi:2015cka}
S.~H. Hendi and R.~Naderi, \emph{{Geometrothermodynamics of black holes in
  Lovelock gravity with a nonlinear electrodynamics}},
  \href{https://doi.org/10.1103/PhysRevD.91.024007}{\emph{Phys. Rev. D}
  {\bfseries 91} (2015) 024007}
  [\href{https://arxiv.org/abs/1510.06269}{{\ttfamily 1510.06269}}].

\bibitem{Dolan:2015xta}
B.~P. Dolan, \emph{{Intrinsic curvature of thermodynamic potentials for black
  holes with critical points}},
  \href{https://doi.org/10.1103/PhysRevD.92.044013}{\emph{Phys. Rev. D}
  {\bfseries 92} (2015) 044013}
  [\href{https://arxiv.org/abs/1504.02951}{{\ttfamily 1504.02951}}].

\bibitem{Mansoori:2016jer}
S.~A.~H. Mansoori, B.~Mirza and E.~Sharifian, \emph{{Extrinsic and intrinsic
  curvatures in thermodynamic geometry}},
  \href{https://doi.org/10.1016/j.physletb.2016.05.096}{\emph{Phys. Lett. B}
  {\bfseries 759} (2016) 298}
  [\href{https://arxiv.org/abs/1602.03066}{{\ttfamily 1602.03066}}].

\bibitem{HosseiniMansoori:2019jcs}
S.~A. Hosseini~Mansoori and B.~Mirza, \emph{{Geometrothermodynamics as a
  singular conformal thermodynamic geometry}},
  \href{https://doi.org/10.1016/j.physletb.2019.135040}{\emph{Phys. Lett. B}
  {\bfseries 799} (2019) 135040}
  [\href{https://arxiv.org/abs/1905.01733}{{\ttfamily 1905.01733}}].

\bibitem{Banerjee:2010da}
R.~Banerjee, S.~Ghosh and D.~Roychowdhury, \emph{{New type of phase transition
  in Reissner Nordström--AdS black hole and its thermodynamic geometry}},
  \href{https://doi.org/10.1016/j.physletb.2010.12.010}{\emph{Phys. Lett. B}
  {\bfseries 696} (2011) 156}
  [\href{https://arxiv.org/abs/1008.2644}{{\ttfamily 1008.2644}}].

\bibitem{Banerjee:2010bx}
R.~Banerjee, S.~K. Modak and S.~Samanta, \emph{{Second Order Phase Transition
  and Thermodynamic Geometry in Kerr-AdS Black Hole}},
  \href{https://doi.org/10.1103/PhysRevD.84.064024}{\emph{Phys. Rev. D}
  {\bfseries 84} (2011) 064024}
  [\href{https://arxiv.org/abs/1005.4832}{{\ttfamily 1005.4832}}].

\bibitem{Wei2015}
S.-W. Wei and Y.-X. Liu, \emph{{Insight into the Microscopic Structure of an
  AdS Black Hole from a Thermodynamical Phase Transition}},
  \href{https://doi.org/10.1103/PhysRevLett.116.169903,
  10.1103/PhysRevLett.115.111302}{\emph{Phys. Rev. Lett.} {\bfseries 115}
  (2015) 111302} [\href{https://arxiv.org/abs/1502.00386}{{\ttfamily
  1502.00386}}].

\bibitem{Guo2019}
X.-Y. Guo, H.-F. Li, L.-C. Zhang and R.~Zhao, \emph{{Microstructure and
  continuous phase transition of a Reissner-Nordstrom-AdS black hole}},
  \href{https://doi.org/10.1103/PhysRevD.100.064036}{\emph{Phys. Rev.}
  {\bfseries D100} (2019) 064036}
  [\href{https://arxiv.org/abs/1901.04703}{{\ttfamily 1901.04703}}].

\bibitem{Du2019}
Y.-Z. Du, R.~Zhao and L.-C. Zhang, \emph{{Microstructure and Continuous Phase
  Transition of the Gauss-Bonnet AdS Black Hole}},
  \href{https://arxiv.org/abs/1901.07932}{{\ttfamily 1901.07932}}.

\bibitem{Dehyadegari2017}
A.~Dehyadegari, A.~Sheykhi and A.~Montakhab, \emph{{Critical behavior and
  microscopic structure of charged AdS black holes via an alternative phase
  space}}, \href{https://doi.org/10.1016/j.physletb.2017.02.064}{\emph{Phys.
  Lett.} {\bfseries B768} (2017) 235}
  [\href{https://arxiv.org/abs/1607.05333}{{\ttfamily 1607.05333}}].

\bibitem{Chabab2018}
M.~Chabab, H.~El~Moumni, S.~Iraoui, K.~Masmar and S.~Zhizeh, \emph{{More
  Insight into Microscopic Properties of RN-AdS Black Hole Surrounded by
  Quintessence via an Alternative Extended Phase Space}},
  \href{https://doi.org/10.1142/S0219887818501712}{\emph{Int. J. Geom. Meth.
  Mod. Phys.} {\bfseries 15} (2018) 1850171}
  [\href{https://arxiv.org/abs/1704.07720}{{\ttfamily 1704.07720}}].

\bibitem{Deng2017}
G.-M. Deng and Y.-C. Huang, \emph{{$Q$-$\Phi$ criticality and microstructure of
  charged AdS black holes in $f(R)$ gravity}},
  \href{https://doi.org/10.1142/S0217751X17502049}{\emph{Int. J. Mod. Phys.}
  {\bfseries A32} (2017) 1750204}
  [\href{https://arxiv.org/abs/1705.04923}{{\ttfamily 1705.04923}}].

\bibitem{Zangeneh2017}
M.~Kord~Zangeneh, A.~Dehyadegari, A.~Sheykhi and R.~B. Mann, \emph{{Microscopic
  Origin of Black Hole Reentrant Phase Transitions}},
  \href{https://doi.org/10.1103/PhysRevD.97.084054}{\emph{Phys. Rev.}
  {\bfseries D97} (2018) 084054}
  [\href{https://arxiv.org/abs/1709.04432}{{\ttfamily 1709.04432}}].

\bibitem{Miao2017}
Y.-G. Miao and Z.-M. Xu, \emph{{On thermal molecular potential among
  micromolecules in charged AdS black holes}},
  \href{https://doi.org/10.1103/PhysRevD.98.044001}{\emph{Phys. Rev.}
  {\bfseries D98} (2018) 044001}
  [\href{https://arxiv.org/abs/1712.00545}{{\ttfamily 1712.00545}}].

\bibitem{Miao2019a}
Y.-G. Miao and Z.-M. Xu, \emph{{Microscopic structures and thermal stability of
  black holes conformally coupled to scalar fields in five dimensions}},
  \href{https://doi.org/10.1016/j.nuclphysb.2019.03.015}{\emph{Nucl. Phys.}
  {\bfseries B942} (2019) 205}
  [\href{https://arxiv.org/abs/1711.01757}{{\ttfamily 1711.01757}}].

\bibitem{Chen2019}
Y.~Chen, H.~Li and S.-J. Zhang, \emph{{Microscopic explanation for black hole
  phase transitions via Ruppeiner geometry: Two competing factors–the
  temperature and repulsive interaction among BH molecules}},
  \href{https://doi.org/10.1016/j.nuclphysb.2019.114752}{\emph{Nucl. Phys.}
  {\bfseries B948} (2019) 114752}
  [\href{https://arxiv.org/abs/1812.11765}{{\ttfamily 1812.11765}}].

\bibitem{Xu:2019nnp}
Z.-M. Xu, B.~Wu and W.-L. Yang, \emph{{The fine micro-thermal structures for
  the Reissner-Nordstr\"{o}m black hole}},
  \href{https://arxiv.org/abs/1910.03378}{{\ttfamily 1910.03378}}.

\bibitem{Kumara:2019xgt}
A.~N. Kumara, C.~L.~A. Rizwan, D.~Vaid and K.~M. Ajith, \emph{{Critical
  Behaviour and Microscopic Structure of Charged AdS Black Hole with a Global
  Monopole in Extended and Alternate Phase Spaces}},
  \href{https://arxiv.org/abs/1906.11550}{{\ttfamily 1906.11550}}.

\bibitem{Kumara:2020mvo}
A.~N. Kumara, C.~L.~A. Rizwan, K.~Hegde, A.~K. M. and M.~S. Ali,
  \emph{{Microstructure and continuous phase transition of a regular Hayward
  black hole in anti-de Sitter spacetime}},
  \href{https://arxiv.org/abs/2003.00889}{{\ttfamily 2003.00889}}.

\bibitem{Wei2019a}
S.-W. Wei, Y.-X. Liu and R.~B. Mann, \emph{{Repulsive Interactions and
  Universal Properties of Charged Anti–de Sitter Black Hole
  Microstructures}},
  \href{https://doi.org/10.1103/PhysRevLett.123.071103}{\emph{Phys. Rev. Lett.}
  {\bfseries 123} (2019) 071103}
  [\href{https://arxiv.org/abs/1906.10840}{{\ttfamily 1906.10840}}].

\bibitem{Wei2019b}
S.-W. Wei, Y.-X. Liu and R.~B. Mann, \emph{{Ruppeiner Geometry, Phase
  Transitions, and the Microstructure of Charged AdS Black Holes}},
  \href{https://doi.org/10.1103/PhysRevD.100.124033}{\emph{Phys. Rev.}
  {\bfseries D100} (2019) 124033}
  [\href{https://arxiv.org/abs/1909.03887}{{\ttfamily 1909.03887}}].

\bibitem{Wei:2019ctz}
S.-W. Wei and Y.-X. Liu, \emph{{Intriguing microstructures of five-dimensional
  neutral Gauss-Bonnet AdS black hole}},
  \href{https://doi.org/10.1016/j.physletb.2020.135287}{\emph{Phys. Lett.}
  {\bfseries B803} (2020) 135287}
  [\href{https://arxiv.org/abs/1910.04528}{{\ttfamily 1910.04528}}].

\bibitem{Kumara:2020ucr}
A.~Naveena~Kumara, C.~A. Rizwan, K.~Hegde and A.~K. M, \emph{{Repulsive
  Interactions in the Microstructure of Regular Hayward Black Hole in Anti-de
  Sitter Spacetime}},
  \href{https://doi.org/10.1016/j.physletb.2020.135556}{\emph{Phys. Lett.}
  {\bfseries B807} (2020) 135556}
  [\href{https://arxiv.org/abs/2003.10175}{{\ttfamily 2003.10175}}].

\bibitem{Wei:2020poh}
S.-W. Wei and Y.-X. Liu, \emph{{Extended thermodynamics and microstructures of
  four-dimensional charged Gauss-Bonnet black hole in AdS space}},
  \href{https://doi.org/10.1103/PhysRevD.101.104018}{\emph{Phys. Rev. D}
  {\bfseries 101} (2020) 104018}
  [\href{https://arxiv.org/abs/2003.14275}{{\ttfamily 2003.14275}}].

\bibitem{Wu:2020fij}
B.~Wu, C.~Wang, Z.-M. Xu and W.-L. Yang, \emph{{Ruppeiner geometry and
  thermodynamic phase transition of the black hole in massive gravity}},
  \href{https://arxiv.org/abs/2006.09021}{{\ttfamily 2006.09021}}.

\bibitem{Xu:2019gqm}
Z.-M. Xu, B.~Wu and W.-L. Yang, \emph{{Ruppeiner thermodynamic geometry for the
  Schwarzschild-AdS black hole}},
  \href{https://doi.org/10.1103/PhysRevD.101.024018}{\emph{Phys. Rev. D}
  {\bfseries 101} (2020) 024018}
  [\href{https://arxiv.org/abs/1910.12182}{{\ttfamily 1910.12182}}].

\bibitem{Ghosh:2019pwy}
A.~Ghosh and C.~Bhamidipati, \emph{{Thermodynamic geometry for charged
  Gauss-Bonnet black holes in AdS spacetimes}},
  \href{https://doi.org/10.1103/PhysRevD.101.046005}{\emph{Phys. Rev.}
  {\bfseries D101} (2020) 046005}
  [\href{https://arxiv.org/abs/1911.06280}{{\ttfamily 1911.06280}}].

\bibitem{Ghosh:2020kba}
A.~Ghosh and C.~Bhamidipati, \emph{{Thermodynamic geometry and interacting
  microstructures of BTZ black holes}},
  \href{https://arxiv.org/abs/2001.10510}{{\ttfamily 2001.10510}}.

\bibitem{Yerra:2020oph}
P.~K. Yerra and C.~Bhamidipati, \emph{{Ruppeiner Geometry, Phase Transitions
  and Microstructures of Black Holes in Massive Gravity}},
  \href{https://arxiv.org/abs/2006.07775}{{\ttfamily 2006.07775}}.

\bibitem{Dehyadegari:2020ebz}
A.~Dehyadegari, A.~Sheykhi and S.-W. Wei, \emph{{Microstructure of charged AdS
  black hole via $P-V$ criticality}},
  \href{https://arxiv.org/abs/2006.12265}{{\ttfamily 2006.12265}}.

\bibitem{Born:1934gh}
M.~Born and L.~Infeld, \emph{{Foundations of the new field theory}},
  \href{https://doi.org/10.1098/rspa.1934.0059}{\emph{Proc. Roy. Soc. Lond. A}
  {\bfseries A144} (1934) 425}.

\bibitem{Gibbons:2001gy}
G.~W. Gibbons, \emph{{Aspects of Born-Infeld theory and string / M theory}},
  \href{https://doi.org/10.1063/1.1419338}{\emph{AIP Conf. Proc.} {\bfseries
  589} (2001) 324} [\href{https://arxiv.org/abs/hep-th/0106059}{{\ttfamily
  hep-th/0106059}}].

\bibitem{Fernando:2003tz}
S.~Fernando and D.~Krug, \emph{{Charged black hole solutions in
  Einstein-Born-Infeld gravity with a cosmological constant}},
  \href{https://doi.org/10.1023/A:1021315214180}{\emph{Gen. Rel. Grav.}
  {\bfseries 35} (2003) 129}
  [\href{https://arxiv.org/abs/hep-th/0306120}{{\ttfamily hep-th/0306120}}].

\bibitem{Dey:2004yt}
T.~K. Dey, \emph{{Born-Infeld black holes in the presence of a cosmological
  constant}}, \href{https://doi.org/10.1016/j.physletb.2004.06.047}{\emph{Phys.
  Lett. B} {\bfseries 595} (2004) 484}
  [\href{https://arxiv.org/abs/hep-th/0406169}{{\ttfamily hep-th/0406169}}].

\bibitem{Cai:2004eh}
R.-G. Cai, D.-W. Pang and A.~Wang, \emph{{Born-Infeld black holes in (A)dS
  spaces}}, \href{https://doi.org/10.1103/PhysRevD.70.124034}{\emph{Phys. Rev.
  D} {\bfseries 70} (2004) 124034}
  [\href{https://arxiv.org/abs/hep-th/0410158}{{\ttfamily hep-th/0410158}}].

\end{thebibliography}\endgroup

\end{document}